\documentclass[a4paper,twoside,final,fleqn,usenatbib]{mnras}

\usepackage{newtxtext,newtxmath}

\usepackage[T1]{fontenc}

\DeclareRobustCommand{\VAN}[3]{#2}
\let\VANthebibliography\thebibliography
\def\thebibliography{\DeclareRobustCommand{\VAN}[3]{##3}\VANthebibliography}

\newcommand{\mz}[1]{{#1}}

\usepackage{graphicx}	
\usepackage{amsmath}
\usepackage{amssymb}	
\usepackage{caption}
\usepackage{rotating}
\newcommand{\gaia}{\textit{Gaia}}
\newcommand{\revi}[1]{{#1}}
\newcommand{\revitwo}[1]{{#1}}

\title[]{Dust extinction map of the Galactic plane based on the UKIDSS survey data}

\author[M. Zhang et al.]{
M. Zhang$^{1}$\thanks{E-mail: miaomiao@pmo.ac.cn},
J. Kainulainen$^{2}$, 
H. Zhao$^{1}$,
Y. Su$^{1}$,
M. Fang$^{1}$,
Y. Ma$^{1}$,
Z. Chen$^{1}$, and 
Z. Jiang$^{1}$
\\
$^{1}$Purple Mountain Observatory, and Key Laboratory for Radio Astronomy, Chinese Academy of Sciences, Nanjing 210023, China\\
$^{2}$Chalmers University of Technology, Department of Space, Earth and Environment, SE-412 93 Gothenburg, Sweden
}

\date{Accepted XXX. Received YYY; in original form ZZZ}

\pubyear{2024}

\begin{document}
\label{firstpage}
\pagerange{\pageref{firstpage}--\pageref{lastpage}}
\maketitle

\begin{abstract}
Dust plays a critical role in the study of the interstellar medium (ISM). \revi{Extinction maps derived from optical surveys often fail to capture regions with high column density due to the limited photometric depth in optical wavelengths. }
\revi{To address these limitations, we developed the XPNICER method based on near-infrared (NIR) photometric survey data. This method combines the previously established PNICER and Xpercentile techniques, enabling effective mitigation of foreground contamination and improved handling of complex dust structures in the Galactic plane, which thus can provide more accurate extinction estimates, particularly in highly obscured regions.} By applying XPNICER to the Galactic Plane Survey from the UKIRT Infrared Deep Sky Survey, we have generated a series of two-dimensional (2D) dust extinction maps that span roughly $\sim$1800 deg$^2$ of the Galactic plane (0\degr~$\lesssim l \lesssim$~110\degr~and 140\degr~$\lesssim l \lesssim$~232\degr, $|b|\lesssim$~5\degr). These maps, with spatial resolutions between 30\arcsec~and \revi{300}\arcsec, can trace extinction up to $A_V\sim$~30$-$40\,mag. This new approach offers higher spatial resolution and better detection of high-extinction regions compared to previous large-scale dust-based maps of the Galactic plane, providing an independent and complementary measure of dust column densities.
\end{abstract}



\section{INTRODUCTION}

Dust plays a crucial role in the interstellar medium (ISM), acting as an important tracer for Galactic ISM and offering insights into its physical properties. The spatial distribution of Galactic dust has been mapped using various datasets. \revi{Dust emission is a widely used method for mapping dust distribution on large scales. It has been used to construct several widely referenced all-sky dust maps that play an important role in many areas of astrophysical research \citep{sfd1998,planck-dust-2014}. However, converting thermal dust emission into dust column density involves assumptions about dust emissivity and the temperature distribution along the line of sight. These assumptions introduce significant uncertainties due to limited knowledge of the physical properties of interstellar dust \citep{ossenkopf1994,draine2009,padoan-pp6}.}

\revi{Another} commonly used method involves measuring reddening toward many stars and treating them as individual samples of a continuous dust distribution. \revi{The extinction affecting a single star can be estimated by comparing its observed spectral energy distribution (SED) with its intrinsic SED. Although this approach ideally requires detailed spectroscopic data for accurate intrinsic characterization, it has the advantage of being independent of assumptions regarding dust temperature and physical properties. Consequently, this approach} avoids the stringent assumptions about dust properties, which makes it a valuable complementary technique for tracing column density compared to methods based on thermal dust emission or gas tracers like $^{13}$CO \citep{draine2009,goodman2009,padoan-pp6}.
 
Many optical studies have focused on determining stellar parameters, distances, and reddening by using wide-field optical surveys and stellar evolutionary models \citep{green2019,bai2020,starhorse2020,stellarpar2023}. These studies generate catalogs that facilitate mapping of the three-dimensional (3D) distribution of dust \citep{rezaei2018,chen2019,green2019,lallement2019,leike2020,eden2023,eden2024}. However, dust maps from optical surveys struggle to capture high column density regions, which are essential for understanding star formation \citep{gao2004,lada2010,mypub2019}. Since extinction decreases with longer wavelengths, near-infrared (NIR) data offers a broader dynamic range for measuring dust column density \citep{lada1994,dobashi2011,juvela16}. However, the lack of extensive NIR spectroscopic and astrometric surveys makes it challenging to obtain accurate stellar parameters. NIR extinction is usually determined by comparing observed NIR colors with average intrinsic ones, benefiting from the low variation in intrinsic colors and extinction laws at NIR wavelengths \citep{cardelli1989,nicer2001,wangjiang2014,meingast2018,wangchen2019,butler2024}.

Various 2D extinction mapping methods, including NICE \citep{lada1994}, NICER \citep{nicer2001}, PNICER \citep{pnicer2017}, \revi{XNICER \citep{xnicer2018}}, and Xpercentile \citep{dobashi2008}, have been developed with distinct approaches. These techniques are commonly used to study column density structures in nearby star-forming regions \citep[e.g.,][]{kai2006,kai2007,kai2009,lombardi06,lombardi08ext,lombardi10,alves14,spilker2021}. However, they are rarely applied to Galactic plane studies, as they are based on measuring the reddening of stars behind dust clouds, which becomes difficult in the Galactic plane due to the complex geometry of stars and dust \citep[e.g.,][]{lombardi05, kainulainen11alves,juvela16,vvvextmap}. \revi{Furthermore, the selection of suitable control fields for estimating intrinsic stellar colors becomes increasingly challenging in the Galactic plane, where most regions are affected by variable extinction.}

To address this challenge, we have developed the NIR extinction mapping technique XPNICER \citep{vvvextmap}, building on PNICER \citep{pnicer2017} and Xpercentile \citep{dobashi2008}. XPNICER estimates stellar intrinsic colors using precise extinction data from Gaia, eliminating the need for nearby extinction-free reference fields. It also minimizes foreground contamination by selecting the reddest X percentile stars as background sources. This method is particularly effective in measuring extinction toward distant and/or dense dust clouds behind many foreground stars. We have applied XPNICER to the Vista Variables in the V\'ia L\'actea (VVV) survey photometric data \citep{vvv2010,vvvdaophotmypub}, creating a 2D extinction map covering the entire VVV survey area in the Galactic plane.

In this study, we extended the use of XPNICER to the UKIRT Infrared Deep Sky Survey \citep[UKIDSS;][]{ukidss2007} Galactic Plane Survey \citep[GPS;][]{ukidss-gps}. We have derived 2D extinction maps with spatial resolutions from 30\arcsec~to 180\arcsec, covering the entire UKIDSS/GPS region, with a dynamic range up to $A_V\sim$~30$-$40\,mag. Section \ref{sect:data} outlines the source catalogs, including UKIDSS/GPS and Gaia DR3, while Section \ref{sect:method} reviews the XPNICER technique. Section \ref{sect:results} presents the extinction maps, and Section \ref{sect:discussion} compares them with several previous dust-based maps. The main findings are summarized in Section \ref{sect:summary}.

\section{DATA}\label{sect:data}
The UKIDSS/GPS catalog provides measurements of the observed colors of sources within the Galactic plane. A key step in our XPNICER method involves estimating the intrinsic colors of these sources. To achieve this, we incorporated the \gaia~DR3 dataset, which enables a statistical estimation of the intrinsic colors of UKIDSS/GPS sources by de-reddening their observed colors using precise extinction measurements from \gaia~DR3.

\subsection{The UKIDSS/GPS catalog}\label{sect:ukidss-gps-catalog}
The UKIDSS/GPS survey \citep[][]{ukidss-gps} covers portions of the northern Galactic plane (0$\degr$~$ \lesssim l \lesssim$~110$\degr$ and 140$\degr$~$\lesssim l \lesssim$~232$\degr$; $|b|\lesssim$~5$\degr$), utilizing the $J, H, K$ filters with the UKIRT Wide Field Camera \citep[WFCAM;][]{wfcam}. UKIDSS/GPS provides arcsecond-scale spatial resolution and achieves median 5$\sigma$ depths of $J =$~19.77, $H = $~19.00, and $K = $~18.05\,mag \citep{ukidss2007}. Additional information about the photometric system, calibration, and data processing can be found in \citet{hewett06}, \citet{hodgkin09}, \citet{irwin08}, and \citet{hambly08}. \revi{For this study, we use the point source catalog from the UKIDSS Galactic Plane Survey Data Release 11 Plus (DR11PLUS\footnote{\url{http://wsa.roe.ac.uk/dr11plus_release.html}}), which includes observations collected between May 2005 and December 2013. We selected point sources using an SQL query with the condition $\mathtt{mergedClass} = -1$ or $-2$, following the recommendation of \citet{ukidss-gps}. To ensure data reliability, we further filtered the sources by requiring $\mathtt{ppErrbits} < 256$, effectively removing detections with known issues. In addition, we applied a photometric uncertainty threshold of 0.35\,mag in the $J$, $H$, and $K_s$ bands to remove detections with unreasonably large errors.} To correct for saturation, sources brighter than $J < $~13.25, $H < $~12.75, or $K <$~12.0\,mag were replaced with photometric data from the Two Micron All Sky Survey \citep[2MASS;][]{2mass2006} point source catalog. \revi{As a result, we obtained approximately 426 million NIR sources within the UKIDSS/GPS survey region, with a median source number density of about 30 arcmin$^{-2}$. This is nearly twice the density of optical point sources detected in the Panoramic Survey Telescope and Rapid Response System Telescope \#1 \citep[Pan-STARRS1;][]{pan-starrs2016}. A detailed comparison of source number densities between the UKIDSS/GPS and Pan-STARRS1 surveys is provided in Appendix~\ref{ap1}.}

\subsection{\gaia~DR3}\label{sect:gaiadr3-catalog}

 \revi{\gaia~Data Release 3 (DR3\footnote{\url{https://gea.esac.esa.int/archive/}})} is based on observations collected over the first 34 months of the \gaia~mission \citep{gaiamission}. DR3 provides highly precise parallax measurements, proper motions, and consistently derived multi-wavelength photometry for around 1.8 billion sources. Additionally, this release includes a wide range of data products, such as spectroscopic observations, photometric time-series, and numerous astrophysical parameters. For this study, we specifically use the astrophysical parameter catalogs from \gaia~DR3.

The astrophysical data products in \gaia~DR3 were generated using 13 distinct modules, each part of the astrophysical parameters inference system \citep[Apsis,][]{apsis2022,apsis-para2022}. For this study, we specifically use the catalog produced by one of these modules, the General Stellar Parameterizer from Photometry \citep[GSP-Phot,][]{gsp-phot2022}. This catalog was derived from a combination of \gaia's astrometry, photometry, and low-resolution BP/RP spectra. GSP-Phot employed a Bayesian forward-modeling approach to create a homogeneous catalog that includes estimates of stellar parameters, distances, and extinctions for roughly 471 million sources with magnitudes of $G<$~19. For bright sources, the typical uncertainty in extinction ($A_G$) is about 0.06 magnitudes. Further details on \gaia~DR3 are provided in \citet{gaiadr3} and \citet{gaiadr3-validation}.

To exclude spurious astrometric data from the \gaia~DR3 catalog, we apply the approach outlined by \citet{rybizki2022}. This method classifies astrometric solutions into ``good'' and ``bad'' categories, using the \gaia~EDR3 dataset as a training set. A machine learning-based parameter, termed ``astrometric fidelity,'' is introduced to identify spurious sources. Compared to traditional quality checks that rely on parameters like $\mathtt{ruwe}$, the ``astrometric fidelity'' parameter offers a more refined selection of sources with reliable astrometric solutions. Additionally, \citet{rybizki2022} provided a method for evaluating photometric contamination from nearby sources ($\mathtt{norm\_dg}$). The following criteria from \citet{rybizki2022} are therefore applied to filter out sources with unreliable astrometric data or anomalous colors in \gaia~DR3:
\mz{\begin{eqnarray}
\mathtt{fidelity\_v2}&>&0.5\label{eq:fidelity},\\
\mathtt{norm\_dg}=\mathtt{nan} ~&or&~ \mathtt{norm\_dg}<-3\label{eq:normdg}
\end{eqnarray}
}

\revi{Finally, we obtained about 63 million \gaia~DR3 sources with GSP-Phot extinction estimates within the UKIDSS/GPS survey region. After applying the filtering criteria described in Eqs.\ref{eq:fidelity} and \ref{eq:normdg}, about 52 million sources with relatively reliable extinction estimates remain.
It is important to note that these ``reliable" GSP-Phot extinction estimates are not suitable for precise measurements of individual stars, but are more appropriately used for statistical analyses of large stellar samples, such as extinction mapping \citep{stellarparamdr2}. Recent studies have also published extinction estimates based on the \gaia~DR3 database using alternative models and algorithms \citep{starhorse2023, zhaohe2024}. While these results are generally consistent in a statistical sense, systematic differences exist among them.
One well-known limitation of the GSP-Phot parameters is the tendency to overestimate both effective temperature and extinction for sources with visual extinction $A_V \gtrsim 2$ mag, due to the degeneracy between temperature and extinction \citep{gsp-phot2022, zhaohe2024}. This issue introduces a zero-point bias in our extinction maps.  We attempted to correct this zero point offset by comparing our XPNICER extinction estimates with that from an external catalog offered by \citet{stellarpar2023} (see Section~\ref{sect:extmap} for details).} \revitwo{It is important to note that extinction catalogs based on \gaia~XP spectra, such as \citet{stellarpar2023}, offer higher precision but are limited to brighter sources, resulting in sparse coverage, especially in heavily obscured regions. Our choice of the \gaia~DR3 GSP-Phot catalog reflects a trade-off between its comprehensive spatial coverage, which is critical for this work, and the higher precision of the sparser \citet{stellarpar2023}'s catalog.}

\subsection{Combined catalog as input of extinction mapping}
\label{sect:input-catalog}

A combined catalog was produced by merging the UKIDSS/GPS point source catalog (see Sect~\ref{sect:ukidss-gps-catalog}) with the astrophysical parameters catalog from \gaia~DR3 (see Sect~\ref{sect:gaiadr3-catalog}), applying a matching tolerance of 0.5\arcsec. \revi{In this work, we did not account for source proper motions during the cross-matching process, as propagating \gaia~astrometry to the UKIDSS/GPS observational epoch for each individual source is computationally complex. To evaluate the potential impact of this simplification, we compared our directly matched catalog with the external cross-matched catalog XGAPS \citep{xgaps2023}.  We found that the contamination of mismatches did not affect the statistical properties of the NIR intrinsic colors of matched sources. A detailed comparison is presented in Appendix~\ref{ap2}. Based on this assessment, we chose to adopt the merged catalog without applying proper motion corrections.} This merged dataset was subsequently used as input for our extinction mapping process, outlined in Sect~\ref{sect:method}. {In the combined catalog, around $\sim$12\% of NIR sources have \gaia~GSP-Phot extinction estimates.}

\section{METHOD}\label{sect:method}

The comprehensive explanation of the XPNICER method is available in \citet{vvvextmap}. In this section, we present an overview of this technique and its application to the UKIDSS/GPS point source catalog. XPNICER, briefly, combines the PNICER \citep{pnicer2017} and X percentile \citep{dobashi2008} approaches. Specifically, it estimates the extinction for individual stars by comparing their observed colors with statistically inferred intrinsic colors from \gaia~GSP-Phot sources in the same region. Next, stars are grouped into discrete sightlines, and potential background sources along different lines of sight are identified using the X percentile method. These background stars are then used to map the spatial distribution of integrated dust extinction.

\subsection{Extinction estimation toward single star}\label{sect:method-singlestar}

In our input catalog (Sect.\ref{sect:input-catalog}), around 50 million sources have \gaia~DR3 extinction measurements. For each source, extinction data from \gaia~DR3 was supplied as a median value along with associated confidence intervals. To simplify the analysis and manage asymmetrical errors, we transformed these intervals into a symmetric standard deviation, assuming Gaussian error distribution. For UKIDSS/GPS sources without \gaia~DR3 extinction measurements, extinction values were estimated using the PNICER method. Specifically, the UKIDSS/GPS survey area was divided into multiple sub-regions, and extinction for sources lacking \gaia~DR3 estimates ($A_{V,\textrm{PNICER}}^{*~\textrm{control:GaiaDR3}}$)\footnote{For clarity, in the following context, we refer to the extinctions of stars as $A_V^*$ and the extinction values of the extinction map as $A_V$.
} was determined by comparing their observed colors to de-reddened colors of sources with \gaia~DR3 measurements within the same sub-region, following the extinction law from \citet{wangchen2019}. 
\revi{The sizes of sub-regions range from 0.5\degr$\times$0.5\degr~to 4\degr$\times$4\degr, in increments of 0.5\degr. This means that for each star, a series of extinction estimates is obtained based on varying sub-region sizes. The final extinction value for each star is adopted as the median of these estimates. This averaging procedure is designed to mitigate the tile effect caused by any particular sub-region size.}
\citet{wangchen2019} examined the extinction law from optical to mid-infrared wavelengths for a sample of red clump stars identified through stellar parameters provided by the APOGEE survey. They measured relative extinction across numerous passbands from surveys such as \gaia, 2MASS, and WISE, indicating an average extinction law in the NIR that is steeper than the commonly used CCM model \citep{cardelli1989}. The uncertainties in $A_{V,\textrm{PNICER}}^{*~\textrm{control:GaiaDR3}}$ are mainly due to photometric errors and variations in intrinsic colors.

However, the selection of UKIDSS/GPS point sources with \gaia~DR3 extinction estimates as reference stars was biased toward nearby and bright stars, introducing additional uncertainties in the PNICER extinction estimates. To assess these uncertainties, we used the Besan{\c c}on model for Galactic stellar population synthesis \citep{besancon2003}, as suggested by \citet{vvvextmap}. For each 1\degr$\times$1\degr~sub-region, we focused on a 30\arcmin$\times$30\arcmin~central area and extracted pseudo-stars from the Besan{\c c}on model. Using linear extinction relations with distance, we created synthetic color-color diagrams (CCDs) and color-magnitude diagrams (CMDs). The best-fit model was identified by minimizing differences between observed and synthetic star density maps in the color space. Extinctions for UKIDSS/GPS point sources ($A_{V,\textrm{PNICER}}^{*~\textrm{control:Besan{\c c}on}}$) were then calculated using the PNICER method, with pseudo-stars from the optimized Besan{\c c}on model serving as reference stars. The standard deviation ($\sigma_{\Delta A_V^*}$) of the extinction difference $\Delta A_V^{*} = A_{V,\textrm{PNICER}}^{*~\textrm{control:Besan{\c c}on}} - A_{V,\textrm{PNICER}}^{*~\textrm{control:GaiaDR3}}$ quantified the extra uncertainty caused by biased reference stars. To reduce edge effects, we smoothed the $\sigma_{\Delta A_V^*}$ map using a Gaussian kernel. The total uncertainty ($\sigma_{\textrm{total}}$) for $A_V^*$ in UKIDSS/GPS sources without \gaia~DR3 extinction estimates combined the PNICER method error ($\sigma_{\textrm{PNICER}}$) with the bias-induced uncertainty ($\sigma_{\Delta A_V^*}$). \revi{In the above process, we did not account for the systematic offset in $\Delta A_V^{*}$, as it strongly depends on the assumed dust distribution along the line of sight and is difficult to quantify without detailed knowledge of the 3D dust structure. However, this systematic offset does affect the zero point of our extinction map. We attempt to estimate this zero-point offset using a data-driven approach in Sect.~\ref{sect:extmap}.}

\subsection{Extinction mapping with potential background sources}\label{sect:background-and-mapping}

To select background sources and map the integrated Galactic extinction, we employed the "X percentile method" as described by \citet{dobashi2008}. The full application of this approach within XPNICER is detailed by \citet{vvvextmap}; here, we provide a brief overview. The UKIDSS/GPS survey region was divided into square sightline cells. These grid cells varied in size from 30\arcsec~to 180\arcsec, where smaller cells improved spatial resolution at the expense of increased noise, and larger cells reduced noise but sacrificed resolution.

We then sorted the UKIDSS/GPS point sources within each grid cell according to their $A_V^*$ values in ascending order. The $q$-th percentile of $A_V^*$ was labeled as $A_V^*(q)$. By defining the $X_0$-th and $X_1$-th percentiles (0\% $\leq X_0 < X_1 \leq $~100\%) of $A_V^*$ as $A_V^*(X_0)$ and $A_V^*(X_1)$, respectively, we selected the UKIDSS/GPS point sources within each grid cell that fall within $A_V^*(X_0) \leq A_V^* \leq A_V^*(X_1)$ as background sources. Following the approach outlined by \citet{vvvextmap}, we chose $X_0 =$~80\% for background source selection and $X_1 =$~95\% to exclude sources with significant infrared excess. However, for distant and/or dense dust clouds, the $X_0 =$~80\% threshold might not adequately identify true background sources. Therefore, we also used an $X_0 =$~90\% threshold, but only for larger grid cells ($\geqslant$60\arcsec). \revi{We note that adopting a threshold of $X_1 = 95\%$ is relatively conservative. \citet{vvvextmap} estimated the fraction of young stellar objects (YSOs) along various lines of sight based on the YSO sample identified by \citet{mypub2019}, and found that the YSO fraction can reach up to $\sim$1\% in regions of dense gas due to YSO clustering. While using $X_1 = 95\%$ may exclude some reddened background sources, it effectively removes sources with infrared excess.}

Finally, we employed a Gaussian kernel to smooth the $A_V^*$ values of the selected background sources, producing the extinction maps. The full width at half maximum (FWHM) of the Gaussian kernels was set to match the size of the grid cells, while the pixel size of the extinction map was half the FWHM of the kernel. \revi{The Gaussian kernel is defined based on the positions and extinction uncertainties of the background sources. In this work, we did not apply corrections for the bias in extinction measurements caused by small-scale cloud substructures \citep{nicest2009}.}
To estimate the uncertainties of the extinction maps, we used a Monte Carlo approach. This method involved assuming Gaussian error distributions for the $A_V^*$ values of the UKIDSS/GPS point sources. We generated random $A_V^*$ values for each point source to create a simulated UKIDSS/GPS catalog. These simulated sources were then organized into a grid, and background sources were selected within each grid cell using the XPNICER technique described earlier. The extinction maps were produced by applying the Gaussian kernel to the $A_V^*$ values of the simulated background sources. This simulation was repeated 10 times, yielding 10 extinction maps. The average of these maps and their standard deviation were used to determine the final extinction map and its associated uncertainty map.

\section{RESULTS}
\label{sect:results}

\subsection{Dust extinction maps and associated uncertainties}
\label{sect:extmap}

\revi{We generated extinction maps, along with corresponding uncertainty and background source number density maps, for both the inner ($l \sim$ 0\degr–110\degr) and outer ($l \sim$ 140\degr–232\degr) Galactic plane. Two sets of maps were produced: one with spatial resolutions of 30\arcsec, 45\arcsec, 60\arcsec, 90\arcsec, and 120\arcsec~using a percentile configuration of $X_0 = 80\%$ and $X_1 = 95\%$, and another with resolutions of 60\arcsec, 90\arcsec, 120\arcsec, 180\arcsec, 240\arcsec, and 300\arcsec~using $X_0 = 90\%$ and $X_1 = 95\%$.
For clarity, we adopt a shorthand notation for the maps throughout the paper. For example, the extinction map with 90\arcsec~resolution and $X_0 = 80\%$ is denoted as $A_V^{90}(X_0=80)$; its associated uncertainty and background source number density maps are denoted as $\delta A_V^{90}(X_0=80)$ and $N_{\mathrm{bg}}^{90}(X_0=80)$, respectively.}

\begin{figure}
    \centering
    \includegraphics[width=1.0\linewidth]{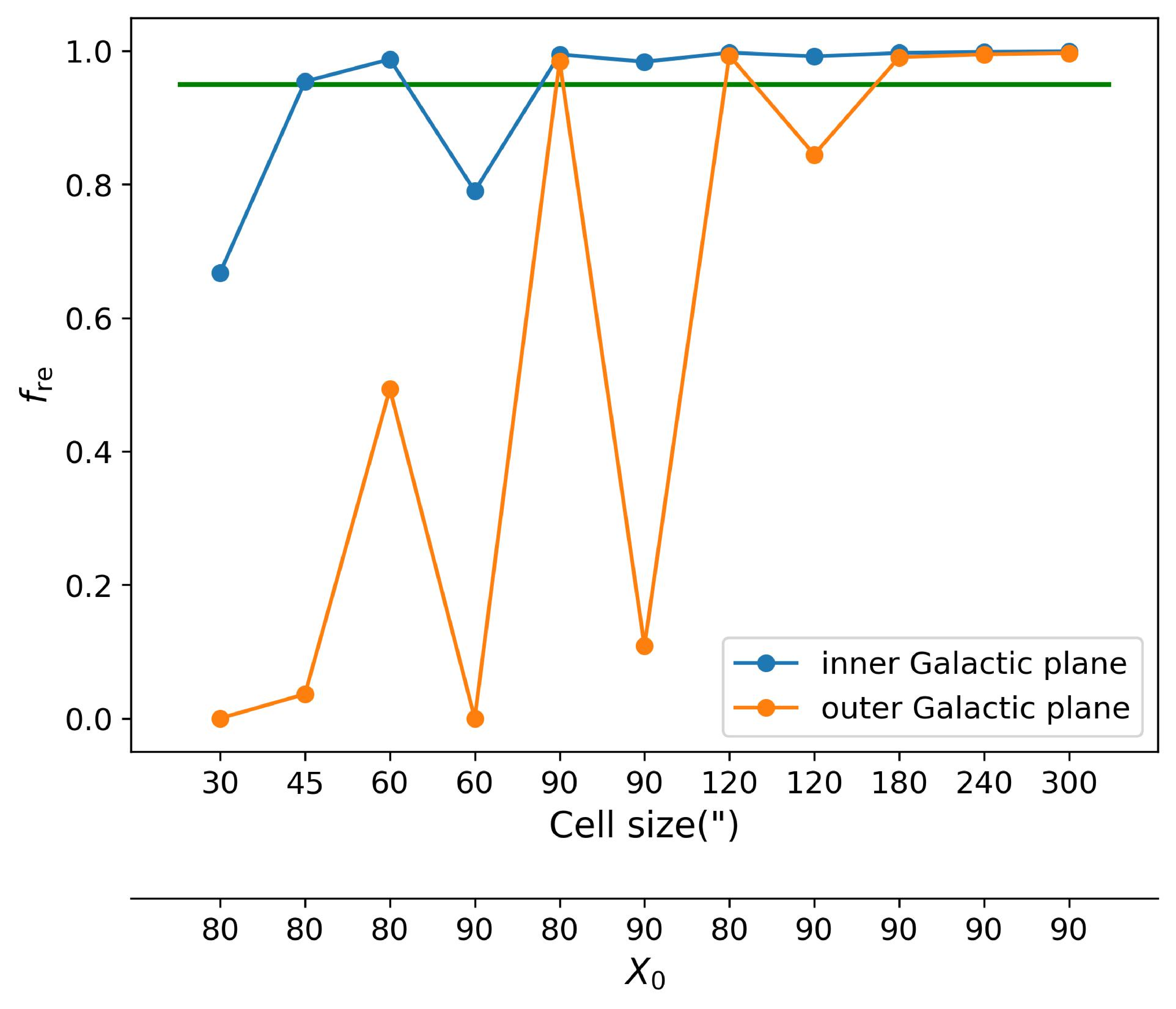}
    \caption{\revi{The fraction of reliable cells relative to the total number of cells, denoted as $f_{\mathrm{re}}$, is presented for various cell sizes and $X_0$ configurations in the extinction maps of both the inner and outer Galactic plane. A green solid line marks the threshold of $f_{\mathrm{re}} = 0.95$. Extinction maps with $f_{\mathrm{re}}$ values below this threshold are considered unreliable, as they contain an insufficient number of background sources in some beams.}}
    \label{fig:beamstar}
\end{figure}

\revi{Cells containing three or more background sources are considered reliable. For each extinction map, we define the reliability fraction, $f_{\mathrm{re}}$, as the ratio of reliable cells to the total number of cells, varying with resolution and $X_0$ configuration. Figure~\ref{fig:beamstar} shows $f_{\mathrm{re}}$ for different extinction maps across various resolutions and $X_0$ values. For example, the inner and outer Galactic plane extinction maps with 60\arcsec\ resolution and $X_0=80\%$ have $f_{\mathrm{re}}$ values of 0.99 and 0.49, respectively. This means that at least three background sources are present in 99\% of the beams in the inner region and 49\% in the outer region for the map $A_V^{60}(X_0=80)$.}

\revi{We propose a rough reliability level of 0.95 for $f_{\mathrm{re}}$, suggesting that the extinction maps $A_V^{30}(X_0=80)$, $A_V^{45}(X_0=80)$, and $A_V^{60}(X_0=90)$ for the inner Galactic plane, extinction maps with resolution of $<$90\arcsec, $A_V^{90}(X_0=90)$, and $A_V^{120}(X_0=90)$ for the outer Galactic plane are unreliable due to the insufficient number of background sources in some beams. Therefore, although we have released all extinction maps with various resolutions and configurations in Sect.~\ref{sect:product}, we recommend that readers only use the reliable maps. Specifically, maps with a resolution of 60\arcsec~or higher--excluding $A_V^{60}(X_0=90)$--should be used for the inner Galactic plane, and maps with a resolution of 90\arcsec~or higher--excluding $A_V^{90}(X_0=90)$ and $A_V^{120}(X_0=90)$--should be used for the outer Galactic plane. It should be noted that the reliability threshold we propose is not a strict or universally applicable limit. Users may define their own reliability criteria based on scientific requirements, using the released background source number density maps as reference.}

\revi{We also released an uncertainty map associated with each extinction map. The uncertainty maps primarily include contributions from observed photometric uncertainties, variations in intrinsic colors, and biases in reference sources. However, as noted by \citet{vvvextmap}, the uncertainty maps do not account for systematic uncertainties, such as zero-point errors and uncertainties due to extinction laws.}

\revi{The zero-point offset in our extinction maps mainly originates from two sources. First, there are systematic biases in the \gaia~GSP-Phot extinction estimates due to the degeneracy between stellar temperature and extinction (see Sect.\ref{sect:gaiadr3-catalog}). Second, systematic errors can arise from the use of reference sources. Since \gaia~extinction measurements are only available for relatively bright stars, using their intrinsic colors to estimate the intrinsic colors of fainter background stars may introduce additional biases (see Sect.\ref{sect:method-singlestar}).
To assess the zero-point offset, we employed a reference catalog that provides extinction estimates ($A_{V,\textrm{Zhang2023}}^*$) and other stellar parameters for $\sim$220 million stars. These parameters were derived from Gaia XP spectra using a data-driven approach \citep{stellarpar2023}.}

\revi{Compared to the GSP-Phot extinction estimates, \citet{stellarpar2023} focused on \gaia~sources with XP spectra and adopted an empirical forward modeling approach that is independent of any stellar evolutionary models. This methodology enables more accurate and robust extinction estimation. To select high-quality extinction measurements from the catalog of \citet{stellarpar2023}, we applied the following criteria:
\begin{align*}
\mathtt{quality}\_\mathtt{flags}&<~8 \\
E&\leq10\\
\sigma_E&\leq0.04\\
\sigma_{\omega}/\omega&<0.1
\end{align*} 
where $E$ and $\omega$ are the extinction and parallax of sources, respectively, while $\sigma_E$ and $\sigma_{\omega}$ are the associated uncertainties. Here $E$ is actually a scalar
proportional to extinction, which can be converted to $A_{V,\mathrm{Zhang2023}}^{*}$ value using a extinction curve. We further required $\omega<$~0.2 mas to select sources that located at large distances of $\gtrsim$~5\,kpc, yielding $\sim$0.5 million sources within the UKIDSS/GPS survey area. For each source with an extinction measurement $A_{V,\mathrm{Zhang2023}}^{*}$, we extracted its corresponding extinction $A_{V,\mathrm{XPNICER}}^*$ from our XPNICER extinction map based on its Galactic coordinates. The extinction difference was defined as $Z_{0}^{*}=A_{V,\textrm{XPNICER}}^*-A_{V,\textrm{Zhang2023}}^*$. We then applied a sigma-clipping algorithm to remove outliers in the $Z_{0}^{*}$ distribution. Finally, the zero-point offset map $Z_0$ was constructed by smoothing $Z_0^*$ with a Gaussian kernel of FWHM=1\degr.} 

\begin{figure*}
    \centering
    \includegraphics[width=1.0\linewidth]{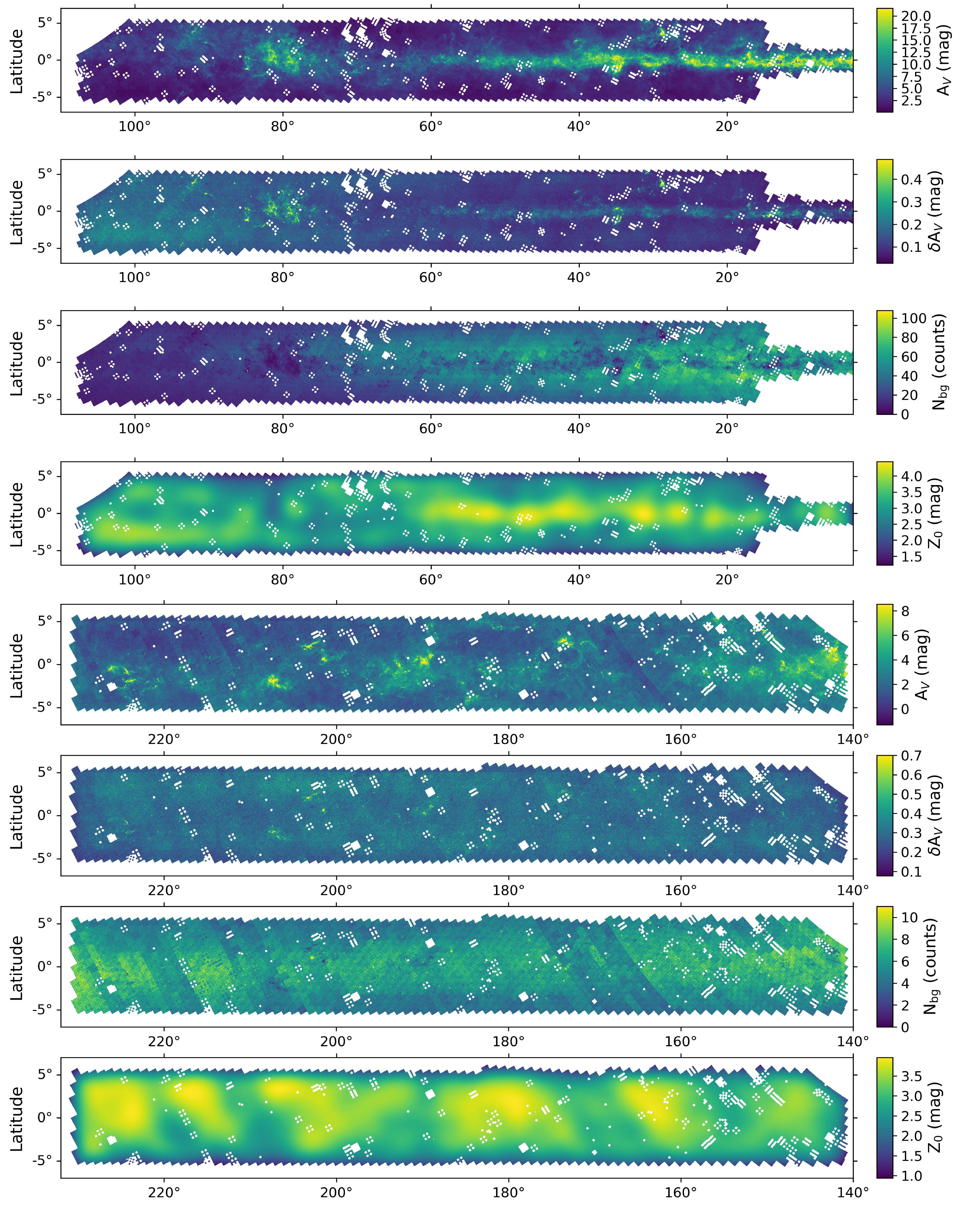}
    \caption{The XPNICER extinction maps, associated uncertainty maps, number density maps of background sources, \revi{and zero-point offset map} obtained using $X_0=$~80\% and $X_1=$~95\% with the spatial resolution of \revi{90\arcsec}~for the Galactic plane area covered by UKIDSS/GPS. \revi{The zero-point offset has been subtracted from the extinction map.}}
    \label{fig:extmap_disk}
\end{figure*}

\revi{Figure~\ref{fig:extmap_disk} shows the $Z_0$ map whose values span from 0.2 to 4.5 mag with a median value of $\sim$3 mag. Obviously, $Z_0$ is the function of the Galactic positions. On average, $Z_0$ of the inner Galactic plane is higher than that of the outer Galactic plane as well as $Z_0$ has higher values in some high extinction regions. We note that the median value ($\sim$3\,mag) of $Z_0$ map is significantly higher than the zero point offset ($\sim$1\,mag) derived by \citet{vvvextmap}. However, \citet{vvvextmap} compared their map with Planck dust map \citep{planck-dust-2014} and found that their XPNICER extinction systematically overestimated $A_V$ value of about 3-4\,mag. Therefore, we believed that \citet{vvvextmap} should underestimate the zero point offset of their extinction map.}

The systematic uncertainties arising from extinction laws have been thoroughly discussed by \citet{vvvextmap}. Generally speaking, our study assumes a universal NIR extinction law for the entire UKIDSS/GPS survey area, as proposed by \citet{wangchen2019}. This law, characterized by $\alpha =$~2.07 $\pm$ 0.03, follows a power-law relation $A_{\lambda} \propto \lambda^{-\alpha}$. Although other studies suggest different $\alpha$ values, from 1.61 to 2.47 \citep{rieke1985,alonso2017,hosek2018,sanders2022}, the variation introduces uncertainties in extinction estimation as discussed by \citet{vvvextmap}. Large-scale investigations generally support a universal NIR extinction law, but specific environments like star forming regions may exhibit different laws \citep{wang2013,wangjiang2014,meingast2018,extlaw2020}. Our chosen $\alpha$ value results in uncertainties of within 3\% for visual extinctions, though systematic errors could reach 20-30\% as suggested by \citet{vvvextmap}.

\revi{Figure~\ref{fig:extmap_disk} also presents the $A_V^{90}(X_0=90)$, $\delta A_V^{90}(X_0=80)$, and $N_{\mathrm{bg}}^{90}(X_0=80)$ for both the inner and outer area of the Galactic plane, as surveyed by UKIDSS/GPS in units of $A_V$. Here we emphasized that the zero-point offset ($Z_0$) has been subtracted from the extinction map ($A_V^{90}(X_0=90)$). In the subsequent context, $A_V^{\mathrm{FWHM}}(X_0)$ denotes the zero-point corrected extinction maps. Figure~\ref{fig:zoomextmap_disk} shows an close view of $A_V^{90}(X_0=80)$, revealing amounts of detailed dust structures in the Galactic plane.}

\revi{Some tile patterns are visible in the extinction, uncertainty, and source number density maps, particularly in the outer Galactic plane regions. These patterns likely result from varying sensitivity across the UKIDSS/GPS survey, as noted by \citet{vvvextmap}, and can be removed by applying a brightness cut to the background sources (see Appendix~\ref{ap:tile-pattern} for details). However, imposing such a cut would significantly reduce the number of background sources, thereby decreasing the resolution and dynamic range of the extinction maps. For this reason, we chose not to correct for the tile pattern effect.}

\subsection{Limiting distance corresponding to integrated extinction}
\label{sect:dlim}

\begin{figure*}
    \centering
    \includegraphics[width=1.0\linewidth]{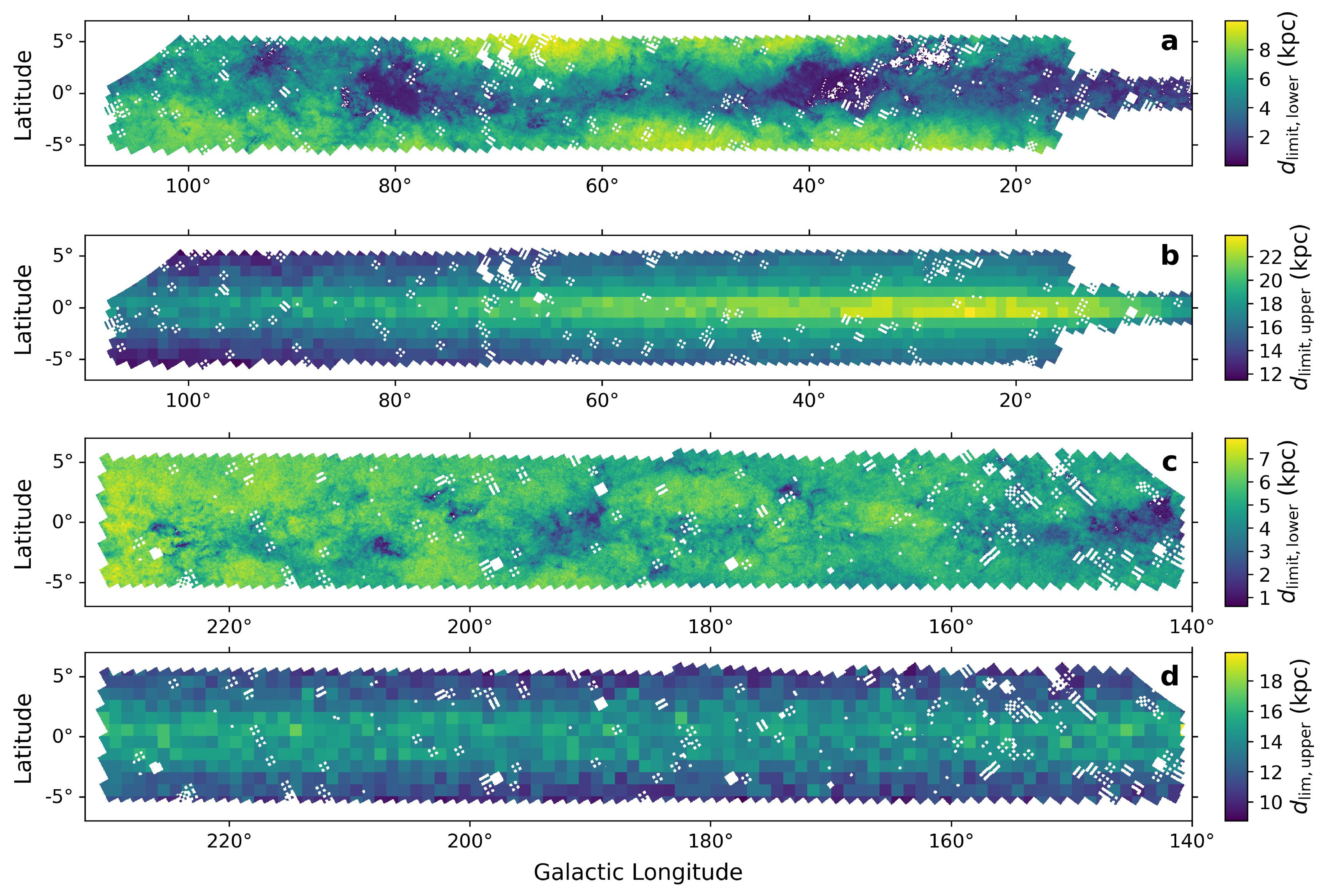}
    \caption{The upper and lower limit of the limiting distance ($d_{\textrm{limit}}$) to which the extinction integrated. Panels \textit{a} and \textit{c} show the lower limit ($d_{\textrm{limit,lower}}$) map of inner and outer Galactic plane, respectively. Panel \textit{b} and \textit{d} present the upper limit ($d_{\textrm{limit,upper}}$) map of inner and outer Galactic plane.}
    \label{fig:dlimrange}
\end{figure*}

By design, our extinction maps are sensitive to extinction integrated to a limiting distance, $d_{\textrm{limit}}$, along the line of sight. This $d_{\textrm{limit}}$ is clearly dependent on Galactic longitude and latitude, as well as the detection limit of the UKIDSS/GPS survey and the 3D dust distribution. We first made rough estimates of the range where $d_{\textrm{limit}}$ might fall. For these estimates, we divided the UKIDSS/GPS coverage area into 1\degr$\times$1\degr sub-regions. In each sub-region, we used the 99.5th percentile of $K$ band magnitudes as the detection limit. Assuming a constant dust distribution profile of $A_V\sim$~0.75\,mag\,kpc$^{-1}$ \citep{lynga1982}, we estimated $d_{\textrm{limit,upper}}$ for each detection limit using the TRILEGAL\footnote{\url{http://stev.oapd.inaf.it/cgi-bin/trilegal}} model \citep{girardi2005}. TRILEGAL is a stellar population synthesis code that simulates the stellar photometry of any Galactic field. Thus, the UKIDSS/GPS survey can detect stars up to $d_{\textrm{limit,upper}}$ along different sight lines, considering only a simple model for the diffuse dust distribution in the Galaxy. Figure~\ref{fig:dlimrange} (b and d) shows the spatial distribution maps of $d_{\textrm{limit,upper}}$, representing the upper limit of $d_{\textrm{limit}}$. We found that $d_{\textrm{limit,upper}}$ varies from $\sim$10 to $\sim$20 kpc among different sightlines.
 
\revi{Since extinction decreases with increasing wavelength, more background stars can be detected at longer wavelengths, such as in NIR. Detecting more distant background stars along a given line of sight implies that the extinction can be integrated out to a larger limiting distance ($d_{\textrm{limit}}$). \citet{lada1994} demonstrated that NIR extinction maps can trace dust column densities with a dynamic range several times greater than that of optical extinction maps, indicating that NIR-based maps typically have higher $d_{\textrm{limit}}$ values. Therefore, we can infer a lower limit for $d_{\textrm{limit}}$ (denoted as $d_{\textrm{limit,lower}}$) by using a published 3D extinction map derived from optical surveys.} \citet{green2019} constructed a 3D dust extinction map of the northern sky ($\textrm{decl.}>-30\degr$) using data from Gaia DR2, PanSTARRS 1 \citep{pan-starrs2016}, and 2MASS. They determined distances, reddenings, and stellar parameters for approximately 799 million stars, mapping the Galactic dust distribution at an angular scale ranging from 3\farcm4 to 13\farcm7. \citet{green2019} also defined the maximum reliable depth to which their map accurately traces dust. Compared to Gaia, PanSTARRS 1 and 2MASS, the UKIDSS/GPS survey detects significantly fainter and more distant stars. Therefore, the maximum reliable depth can be considered a conservative lower limit of $d_{\textrm{limit}}$. Figure~\ref{fig:dlimrange} (a and c) shows the spatial distribution map of $d_{\textrm{limit,lower}}$, equivalent to the maximum reliable depth defined by \citet{green2019}. In the Galactic plane covered by UKIDSS/GPS, $d_{\textrm{limit,lower}}$ ranges from approximately 1 to 10 kpc.

\begin{figure*}
    \centering
    \includegraphics[width=1.0\linewidth]{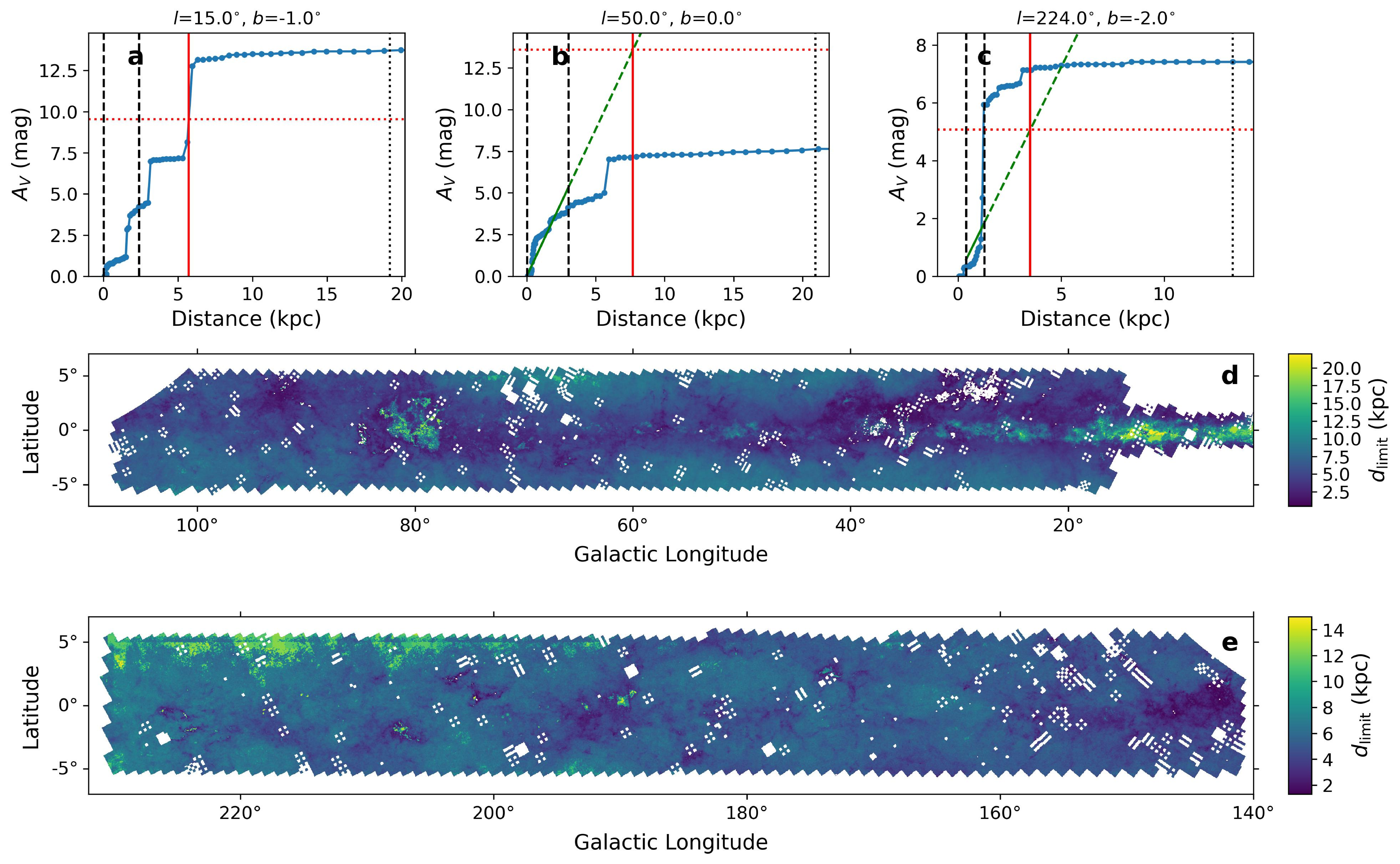}
    \caption{Panels \textit{a}, \textit{b}, and \textit{c} display the visual extinction profiles as functions of distance for three sightlines from \citet{green2019}. The blue points and lines represent these extinction profiles. The minimum and maximum reliable depths ($d_{\textrm{min}}$, $d_{\textrm{max}}$) defined by \citet{green2019} are indicated by black dashed vertical lines. \revi{Panels \textit{a}, \textit{b}, and \textit{c} illustrate the method used to quantify $d_{\mathrm{limit}}$. As described in the main text, we first estimate the lower and upper bounds of $d_{\mathrm{limit}}$, denoted as $d_{\mathrm{limit,lower}}$ and $d_{\mathrm{limit,upper}}$, respectively. The lower limit ($d_{\mathrm{limit,lower}} = d_{\mathrm{max}}$) is indicated by a vertical black dashed line, while the upper limit is shown by a vertical black dotted line. The horizontal red dotted line represents the integrated extinction ($A_{V, \mathrm{XPNICER}}$) from our XPNICER map along each sightline.In panel \textit{a}, the horizontal red dotted line intersects the extinction profile from \citet{green2019} (shown in blue) within the range [$d_{\mathrm{limit,lower}}$, $d_{\mathrm{limit,upper}}$]. The distance at this intersection is taken as $d_{\mathrm{limit}}$ for that sightline and is marked with a vertical red solid line. In panels \textit{b} and \textit{c}, no intersection occurs between $A_{V, \mathrm{XPNICER}}$ and the extinction profiles within the specified range. In these cases, we fit the extinction profile between $d_{\mathrm{min}}$ and $d_{\mathrm{max}}$ (i.e., the section of the blue line bounded by the two vertical black dashed lines) using a linear dust model, as detailed in the main text. The resulting fits are shown as green solid and dashed lines. Where the green dashed line intersects the horizontal red dotted line, we adopt the corresponding distance as $d_{\mathrm{limit}}$, again marked with a vertical red solid line.} Panels \textit{d} and \textit{e} show the final $d_{\textrm{limit}}$ map in the inner and outer Galactic plane, respectively.}
    \label{fig:extprofiles}
\end{figure*}

Although fully quantifying $d_{\textrm{limit}}$ without a detailed 3D model of Galactic dust is not feasible, we provided a rough estimate based on the 3D extinction map by \citet{green2019}. Specifically, \citet{green2019} presented visual extinction profiles as functions of distance for various sightlines. Figure~\ref{fig:extprofiles} (panels a, b, and c) illustrates the relationships between extinction and distance along three sightlines. \citet{green2019} defined the minimum ($d_{\textrm{min}}$) and maximum reliable depths ($d_{\textrm{max}}=d_{\textrm{limit,lower}}$) for each sightline. However, the extinction profile actually extends over the maximum reliable depth and up to 63 kpc, although data beyond such a large distance are sparse. Thus, if our XPNICER extinction value ($A_{V,\textrm{XPNICER}}$) intersects \citet{green2019}'s extinction profile within the distance range [$d_{\textrm{limit,lower}}$, $d_{\textrm{limit,upper}}$], we set the intersection point's distance as $d_{\textrm{limit}}$, as shown in Fig.~\ref{fig:extprofiles}(a). Otherwise, we assumed a linear dust model along each sightline:
\begin{align}
A_{V} = aD\label{eq:dustmodel}
\end{align}
where $A_V$ and $D$ are the visual extinction and distance, respectively. We fit \citet{green2019}'s extinction profile in the range [$d_{\textrm{min}}$, $d_{\textrm{max}}$] along each sightline using Eq.~\ref{eq:dustmodel} with bounded constraints of [$\frac{A_{V,\textrm{XPNICER}}}{d_{\textrm{limit,upper}}}$, $\frac{A_{V,\textrm{XPNICER}}}{d_{\textrm{limit,lower}}}$] for the slope $a$. We then determined $d_{\textrm{limit}}$ using $\frac{A_{V,\textrm{XPNICER}}}{a}$, as shown in Fig.~\ref{fig:extprofiles} (panels b and c). These bounded constraints ensured $d_{\textrm{limit}}$ falls within [$d_{\textrm{limit,lower}}$, $d_{\textrm{limit,upper}}$]. 

Finally, we obtained the $d_{\textrm{limit}}$ maps, shown in Fig.~\ref{fig:extprofiles} (panels d and e) for the inner and outer Galactic plane. We found that $d_{\textrm{limit}}$ ranged from $\sim$2$-$20 kpc. \revi{In some low-extinction regions dominated by diffuse dust, the limiting distance ($d_{\textrm{limit}}$) can reach $\sim$18–20\,kpc, approaching the level of the ``total Galactic dust extinction." Compared to our maps, previous large-scale NIR extinction maps, such as those by \citet{dobashi2011} and \citet{juvela16}, were unable to capture the full extent of extinction, primarily due to their underestimation of the contribution from diffuse dust components \citep{vvvextmap}.
However, it is important to emphasize that the $d_{\textrm{limit}}$ values presented here are rough estimates with large uncertainties. For example, some high-extinction regions near $b=0\degr$ also show $d_{\textrm{limit}}$ values of $\sim$20\,kpc. This does not indicate that our extinction map fully traces the total dust column in these dense areas. Rather, it reflects the complexity of the dust distribution along these lines of sight, which cannot be described by a simple linear model.}

\section{DISCUSSION}\label{sect:discussion}

In this section, we compare our newly developed XPNICER extinction map with several prior dust-based maps. Using VVV survey data, \citet{vvvextmap} produced an XPNICER extinction map of the southern Galactic plane and compared it with ten pre-existing dust-based maps, including the all-sky 2D dust extinction maps \citep{dobashi2011,juvela16}, opacity maps derived from dust emission \citep{sfd1998,planck-dust-2014,ppmap2017}, and 3D dust extinction maps \citep{schultheis2014}. They found overall consistency across all comparisons. Consequently, instead of repeating analyses with numerous maps, we limited our comparison here to three: our previous XPNICER map of the southern Galactic plane \citep{vvvextmap}, the commonly used Planck dust map \citep{planck-dust-2014}, and the 3D extinction map by \citet{green2019}.

The Planck dust map \citep{planck-dust-2014} is chosen as a representative of opacity maps based on dust emission. While the point process mapping (PPMAP) method, applied to the Herschel infrared Galactic Plane (Hi-GAL) survey \citep{ppmap2017}, provides dust maps with a higher spatial resolution ($\sim$12$\arcsec$), it is limited in coverage to Galactic plane latitudes of $|b|\lesssim 1\degr$. Furthermore, \citet{vvvextmap} found that due to inaccuracies in the absolute flux calibration of the Herschel Hi-GAL images, the zero-point of PPMAPs may lack uniformity, which complicates consistent, large-scale comparisons between PPMAPs and other dust maps.

We selected the 3D extinction map by \citet{green2019} for comparison, as it aligns well with our XPNICER map in terms of coverage, spatial resolution, and sensitivity. Although numerous 3D dust extinction maps have been published, some feature low spatial resolutions ($\gtrsim$10$\arcmin$) \citep{marshall2006,hottier2020,dharma2024}, while others with high resolution are limited to tracing integrated dust distributions up to a small distance ($\lesssim$3 kpc) \citep{lallement2019,lallement2022,leike2019,leike2020,vergely2022,eden2023,eden2024}. \citet{chen2019} also created a 3D dust reddening map with about 6$\arcmin$ resolution across the entire Galactic plane, based on \gaia~and 2MASS data. However, \citet{green2019}'s map can trace dust densities to farther distances than \citet{chen2019}'s, benefiting from the deeper photometric data provided by Pan-STARRS 1.

For simplicity, we refer to our XPNICER extinction map as $A_V$(XPNICER), and to the other dust-based maps by \citet{vvvextmap}, \citet{planck-dust-2014}, and \citet{green2019} as $A_V$(Zhang2022), $A_V$(Planck), and $A_V$(Green2019), respectively. It is important to note that the Planck dust map was originally obtained in units of optical depth, specifically $\tau_{\mathrm{353}}$(Planck), where $\tau_{\mathrm{353}}$ represents the optical depth at a frequency of 353 GHz ($\sim$850\,$\mu$m). The color excess $E(B-V)$ map was subsequently derived using a conversion factor from $\tau_{\mathrm{353}}$ to $E(B-V)$, which was estimated based on extinction measurements of extragalactic objects by \citet{planck-dust-2014}. Finally, the $A_V$(Planck) map was obtained by applying the relation between $A_V$ and $E(B-V)$ as suggested by \citet{schlafly2011}.

We present visual comparisons of $A_V$(XPNICER) with the other maps in Sect.~\ref{sect:visualcomp}. Following that, we make quantitative pixel-to-pixel comparisons in Sect.~\ref{sect:compare2vvv}-\ref{sect:compare2green}.

\subsection{Visual comparison}\label{sect:visualcomp}

\begin{figure*}
    \centering
    \includegraphics[width=1.0\linewidth]{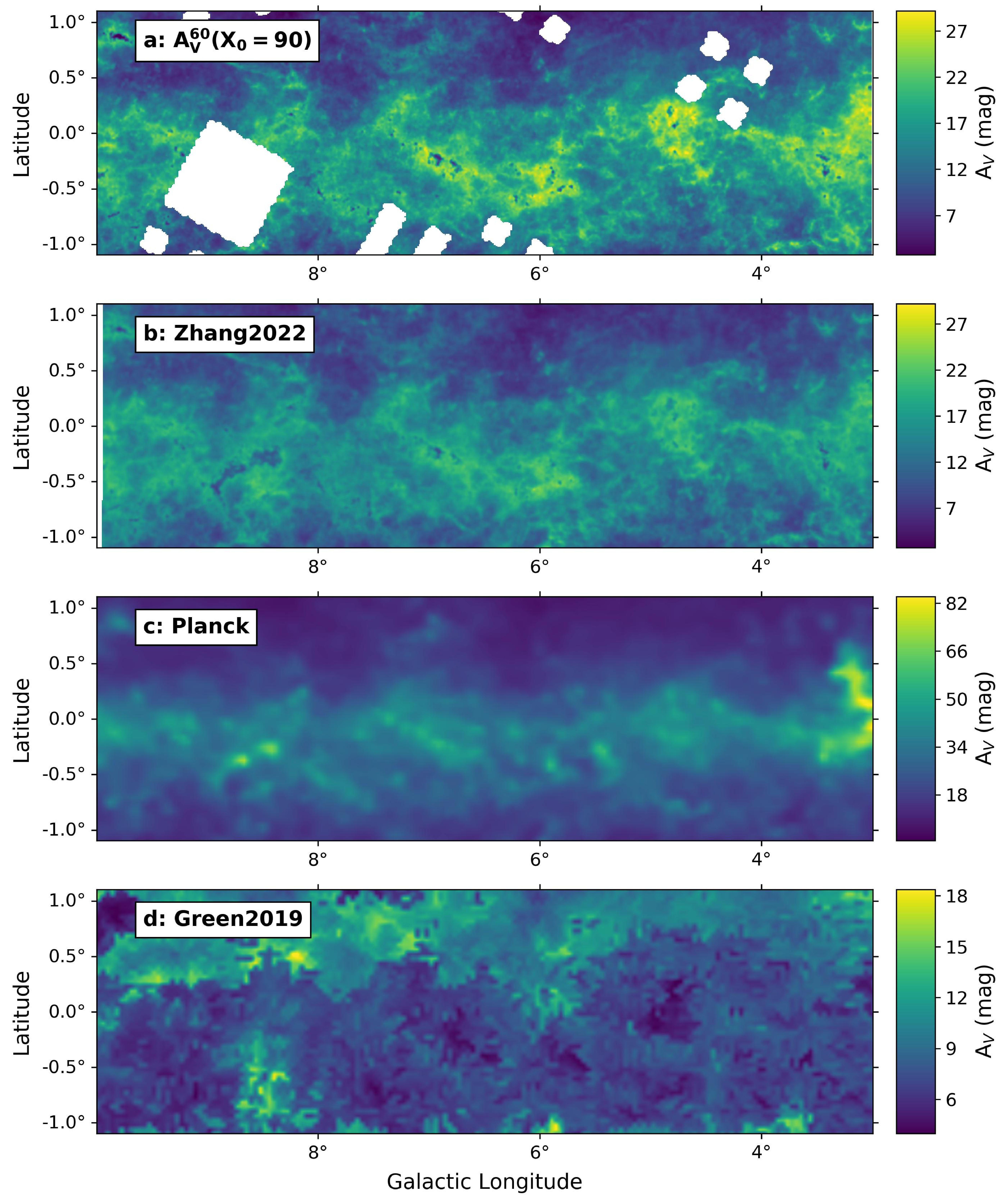}
    \caption{Zoom-in view of a section in the inner Galactic plane of (a): our $A_V^{60}(X_0=90)$ map; (b): XPNICER extinction map obtained by \citet{vvvextmap}. \revi{We also corrected the zero point offset of this map using the method described in Sect.~\ref{sect:extmap}}; (c): Planck dust map \citep{planck-dust-2014}; and (d): integrated 3D extinction map obtained by \citet{green2019}.}
    \label{fig:extmap_zoomin}
\end{figure*}

Figure~\ref{fig:extmap_zoomin} provided a close-up view of a section in the inner Galactic plane across the $A_V$(XPNICER), $A_V$(Zhang2022), $A_V$(Planck), and $A_V$(Green2019) maps. \revi{The $A_V$(Zhang2022) map was also constructed using the XPNICER mapping technique \citep{vvvextmap}. However, its zero-point was not corrected in the original work. Following the method described in Sect.~\ref{sect:extmap}, we computed a zero-point offset map for $A_V$(Zhang2022) and subtracted it to obtain a corrected extinction map. Then $A_V$(Zhang2022) was} compared with $A_V$(XPNICER) under the same configuration of FWHM~$=$~60\arcsec~and $X_0=$~90. This comparison revealed that both maps exhibit nearly identical extinction structures, though $A_V$(Zhang2022) appeared smoother than $A_V$(XPNICER). Actually, the average extinction uncertainties in $A_V$(XPNICER) and $A_V$(Zhang2022) were \revi{0.24} and 0.12 mag, respectively, indicating that $A_V$(Zhang2022) possessed a higher signal-to-noise ratio. This is due to $A_V$(Zhang2022) being derived from a deep photometric catalog obtained by performing PSF photometry on stacked multi-epoch VVV images \citep{vvvdaophotmypub}.

In comparison with $A_V$(Planck), our $A_V$(XPNICER) map under the configuration of \revi{FWHM~$=$~90\arcsec~and $X_0=$~80, i.e., $A_V^{90}(X_0=80)$}, offered higher spatial resolution, enabling the detection of finer structures. Nevertheless, $A_V$(Planck) can effectively trace highly dense regions, with $A_V$ values reaching up to around 90 mag, due to its reliance on far-infrared dust emission, which can be observed even in regions where starlight is significantly obscured. 
When comparing $A_V$(XPNICER), i.e., $A_V^{90}(X_0=80)$, to $A_V$(Green2019), one obvious difference was that $A_V$(XPNICER) revealed more small-scale structures due to its higher resolution. Moreover, the dense regions in $A_V$(XPNICER) often correspond to unreasonably low extinction features in $A_V$(Green2019). This discrepancy arises from the limitations of optical surveys, which become star-limited in areas of high column density. Consequently, compared to extinction maps derived from optical surveys such as $A_V$(Green2019), $A_V$(XPNICER) is better suited for tracing relatively dense dust structures.

\subsection{Detailed comparison with extinction map derived by \citet{vvvextmap}}\label{sect:compare2vvv}

\begin{figure}
\includegraphics[width=1.0\linewidth]{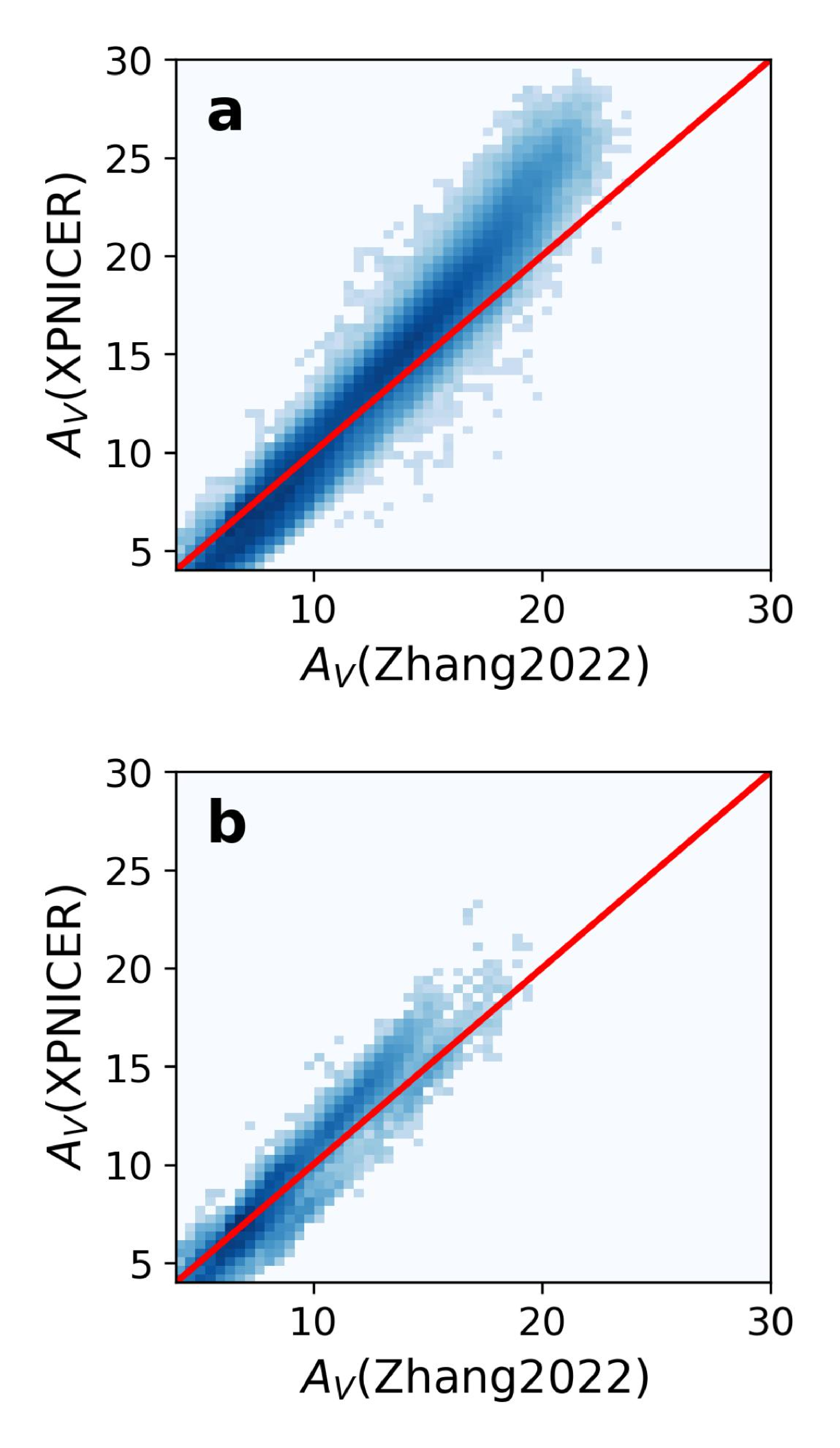}
\caption{\revi{Pixel-by-pixel comparison between our extinction map $A_V$(XPNICER) and the previous XPNICER extinction map $A_V$(Zhang2022) from \citet{vvvextmap} for (\textit{a}): all pixels; and (\textit{b}): reliable pixels with background source number density of $>$~10. The red lines represent the one-to-one correspondence. We have corrected the zero point offset of $A_V$(Zhang2022) using the method described in Sect.~\ref{sect:extmap}.}}
\label{fig:compare2vvv}
\end{figure}

\revi{Figure~\ref{fig:compare2vvv} presents a pixel-by-pixel comparison between $A_V$(XPNICER) and $A_V$(Zhang2022) for the region shown in Fig.~\ref{fig:extmap_zoomin}. Figure~\ref{fig:compare2vvv}a displays the comparison using all pixels from the $A_V$(XPNICER) map, while Fig~\ref{fig:compare2vvv}b includes only the reliable pixels, defined as those with a background source number density greater than 10. We can see that} $A_V$(XPNICER) agrees well with $A_V$(Zhang2022) within the low extinction range of approximately 5$-$15 mag. However, beyond $\sim$15$-$20 mag, notable discrepancies arise, with $A_V$(XPNICER) appearing to overestimate extinction values compared to $A_V$(Zhang2022) in regions with high extinction. 

\revi{The systematic differences between $A_V$(XPNICER) and $A_V$(Zhang2022) could result from the use of different stellar reference catalogs, each introducing its own zero-point offset into the corresponding XPNICER extinction maps. In our study, we used the Gaia DR3 GSP-Phot extinction measurements \citep{gaiadr3gspphot} as the reference, whereas $A_V$(Zhang2022) was based on stellar extinctions from the StarHorse2019 catalog \citep{starhorse2019}. As reported by \citet{gaiadr3gspphot}, there is a known systematic offset between these catalogs: extinctions from StarHorse catalogs are typically higher than those from Gaia DR3 GSP-Phot, especially in regions with high extinction.
To reduce the impact of these zero-point offsets, we applied separate zero-point corrections to each map using a third reference catalog from \citet{stellarpar2023} (see Sect.~\ref{sect:extmap}). Ideally, this calibration should remove systematic biases and bring the corrected maps into agreement across all levels of extinction. In practice, however, this correction is limited by the depth of the \citet{stellarpar2023}'s catalog, which primarily includes stars with extinction values up to approximately $A_V\sim$~10 mag. As a result, the corrections in regions with higher extinction are subject to significant uncertainties due to the lack of suitable calibration stars. These limitations lead to residual systematic differences between the two maps in areas of high extinction.} 

Overall, while $A_V$(XPNICER) shows agreement with $A_V$(Zhang2022) in low-extinction areas, it overestimates extinction values in high-extinction regions due to significant differences between the reference sources used in this study and those in \citet{vvvextmap}.

\subsection{Detailed comparison with Planck dust map}\label{sect:compare2planck}

\citet{planck-dust-2014} developed an all-sky dust model that utilized emission data from both the Planck and IRAS surveys. This model fit the spectral energy distribution (SED) of the emission using a modified blackbody, assuming thermal equilibrium in the optically thin regime. The resulting all-sky map of $\tau_{353}$ was produced at a resolution of 5\arcmin.

\begin{figure}
\includegraphics[width=1.0\linewidth]{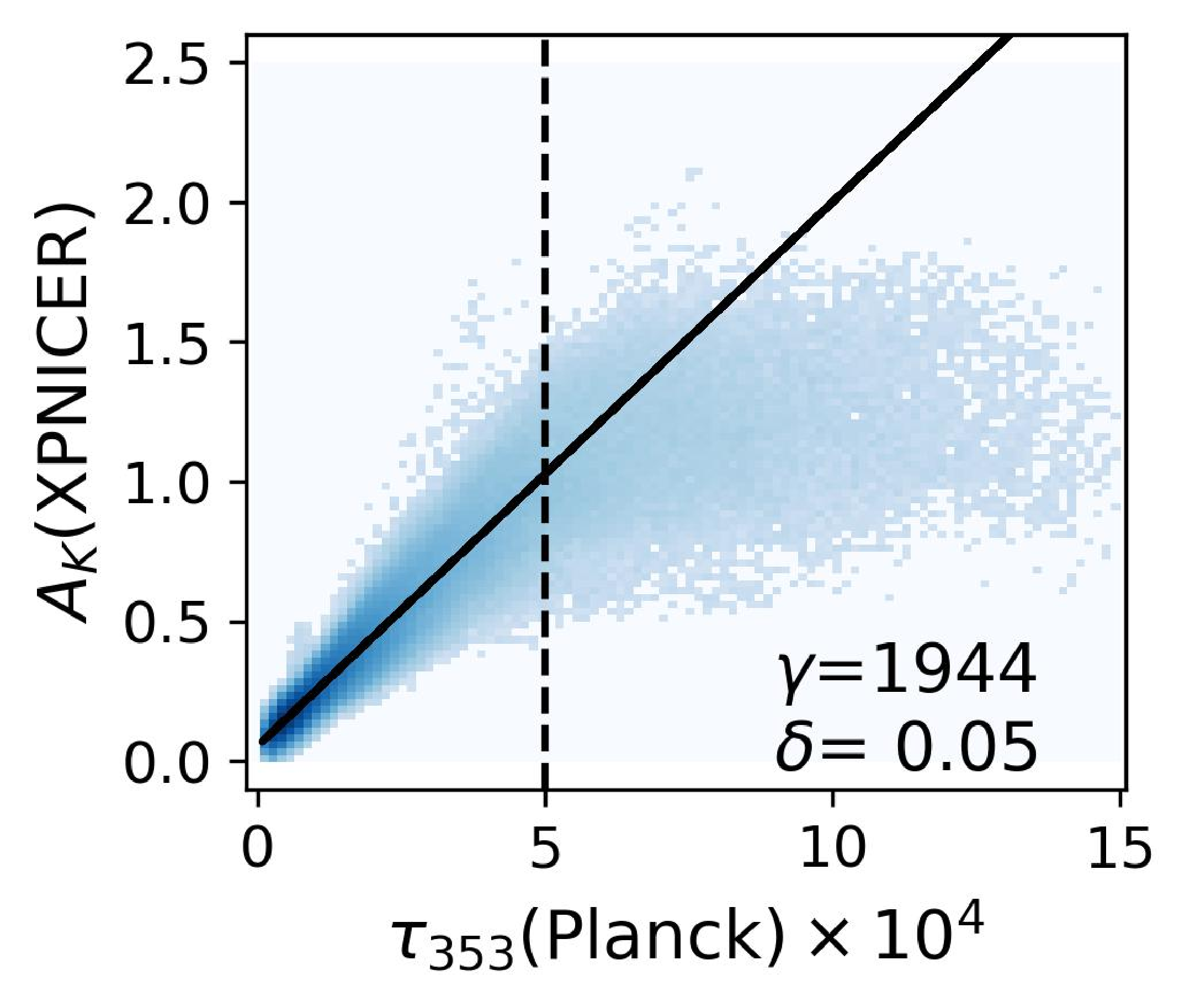}
\caption{Pixel-to-pixel comparison between our XPNICER extinction map and the Planck dust map \citep{planck-dust-2014}. The black solid line represents the linear fit with a slope of $\gamma$ and an intercept of $\delta$. The fitting was constrained to $\tau_{353}<$~$\tau_{\mathrm{cut}}\times$10$^{-4}$, where $\tau_{\mathrm{cut}}=$~5, as indicated by the vertical black dashed line. The fitting parameters are also displayed on the figure.}
\label{fig:compare2planck}
\end{figure}

We retrieved the $\tau_{353}$(Planck) using the Python interface provided by DUSTMAPS \citep{dustmaps}. For consistency, we also converted $A_V$(XPNICER) back to $A_K$ units, denoted as $A_K$(XPNICER), following the reddening law proposed by \citet{wangchen2019}. 
Figure~\ref{fig:compare2planck} displays the pixel-to-pixel comparison between $A_K$(XPNICER) and $\tau_{353}$(Planck) across the entire $A_K$(XPNICER) map. There is an approximate linear relationship for $\tau_{353}\mathrm{(Planck)}\lesssim$~5$\times$10$^{-4}$, but significant discrepancies appear when $\tau_{353}\mathrm{(Planck)}\gtrsim$~5$\times$10$^{-4}$. These discrepancies in dense regions could result from biases in $A_K$(XPNICER) due to limitations in our extinction mapping technique, as noted by \citet{vvvextmap}.

We also applied a linear fit to the relationship between $A_K$(XPNICER) and $\tau_{353}$(Planck), described by the following equation:
\begin{align}\label{eq:tau2ak}
    A_K\mathrm{(XPNICER)} = \gamma \tau_{353}\mathrm{(Planck)} + \delta.
\end{align}
The fitting was constrained to the range where $\tau_{353}$(Planck)~$<$~$\tau_{\mathrm{cut}}\times$10$^{-4}$, with $\tau_{\mathrm{cut}}$ set to 5. \revi{The resulting values for the fitting parameters were $\gamma = 1944$ and $\delta = 0.05$.} Similarly, \citet{vvvextmap} performed a comparison between their XPNICER extinction map and the Planck dust map \citep{planck-dust-2014} using the same linear fitting approach. They obtained fitting parameters of $\gamma_{\mathrm{Zhang2022}} = 2043$ and $\delta_{\mathrm{Zhang2022}} = 0.3$.

\revi{The intercept value of $\delta = 0.05$} from our fit suggests that $A_K$(XPNICER) might systematically overestimate $A_K$ values by about \revi{0.05} in low extinction regions, which corresponds to an $A_V$ value of approximately \revi{0.6\,mag}. \revi{This systematic offset could be attributed to the zero-point uncertainty in our extinction map (see Sect.~\ref{sect:extmap}) and the uncertainty in the Planck dust maps.}

The slope of the linear fit, represented by $\gamma$, is proportional to the ratio of dust opacity to extinction coefficient, as indicated by \citet{lombardi2014}, and is related to dust properties such as composition and grain size distribution \citep{ossenkopf1994}. \citet{vvvextmap} found that their $\gamma_{\mathrm{Zhang2022}} = 2043$ was reasonably within the range of approximately 1000$-$6000, as inferred by previous studies \citep{kramer2003,lombardi2014,zari2016,meingast2018}. \revi{Our similar value of $\gamma = 1944$} should also be consistent with the findings of those prior studies.

\subsection{Detailed comparison with 3D extinction map by \citet{green2019}}\label{sect:compare2green}

We accessed the 3D extinction map presented by \citet{green2019} using the DUSTMAPS Python interface \citep{dustmaps} and integrated it up to a distance of 63 kpc to generate a projected extinction map. This process ultimately provided the integrated 2D extinction map of $A_V$(Green2019).

\begin{figure}
\includegraphics[width=1.0\linewidth]{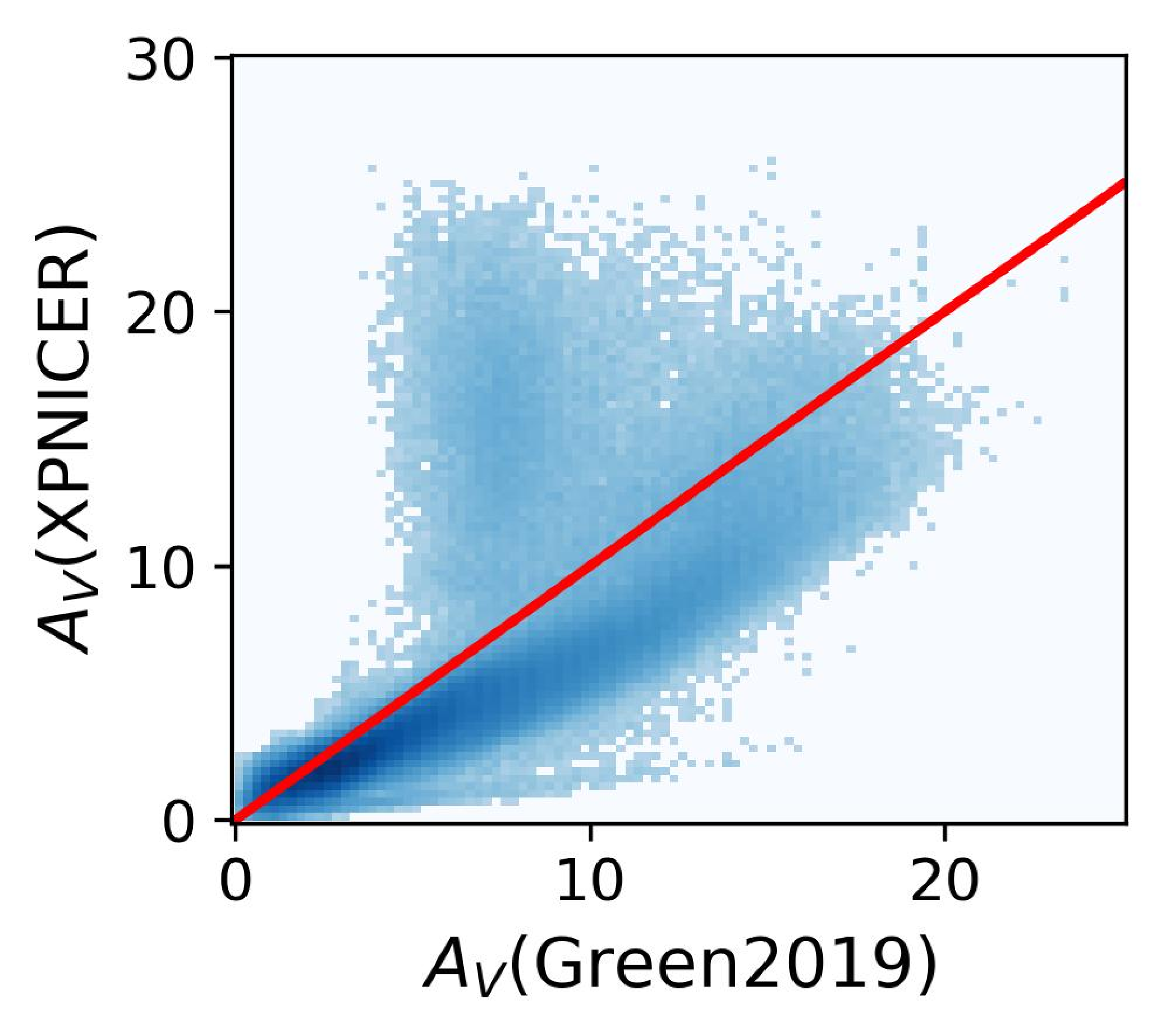}
\caption{Pixel-to-pixel comparison between our XPNICER extinction map and the integrated 3D extinction map by \citet{green2019}. The red line represents the one-to-one relation.}
\label{fig:compare2green}
\end{figure}

Figure~\ref{fig:compare2green} displays the pixel-to-pixel comparison between $A_V$(XPNICER) and $A_V$(Green2019). \revi{While a general correlation is observed between the two maps, $A_V$(XPNICER) also shows a rough agreement with $A_V$(Green2019) in low extinction regions.} However, in high extinction regions ($A_V$(Green2019)$\gtrsim$~5 mag), $A_V$(XPNICER) values are noticeably higher than those in $A_V$(Green2019). This discrepancy likely arises from the limitations of optical surveys, where there are insufficient optical stars to accurately estimate extinction in higher extinction regions, as seen in extinction maps based on optical surveys like \citet{green2019}.

\section{SUMMARY AND CONCLUSIONS}\label{sect:summary}
We have generated 2D dust extinction maps with spatial resolutions ranging from 30\arcsec~to \revi{300}\arcsec~using the UKIDSS/GPS photometric catalog \citep{ukidss-gps}. These maps cover the entire $\sim$1800 deg$^2$ area of the Galactic plane surveyed by UKIDSS/GPS. The maps were produced utilizing the XPNICER technique \citep{vvvextmap}, an advancement of the previous PNICER \citep{pnicer2017} and Xpercentile \citep{dobashi2008} methods. The primary findings and conclusions are summarized as follows:

\begin{itemize}
    \item[1.]{We developed a set of novel dust extinction maps covering the UKIDSS/GPS survey area, employing the XPNICER mapping technique. These maps, with spatial resolutions of 30\arcsec~to \revi{300}\arcsec, are capable of tracing dust extinction up to $A_V\sim$~\revi{35}\,mag. They reflect the total extinction up to distances ranging from $\sim$2 to 20 kpc, which vary significantly depending on the line of sight. We have also made these extinction maps, along with the associated uncertainty and number density maps available at \url{https://doi.org/10.57760/sciencedb.12869}.}
    
    \item[2.]{The typical uncertainty in our XPNICER $A_V$ map is approximately \revi{0.2}\,mag. As suggested by \citet{vvvextmap}, sources of this uncertainty include observed photometric errors, variations in intrinsic colors, and biases in reference stars. Additionally, \revi{the zero-point offset of our XPNICER map is from 0.2$-$4.5\,mag, varying towards the different lines of sight}. The systematic uncertainties arising from different extinction laws and other unclear origins can even reach $\gtrsim$20-30\%.}
    
    \item[3.]{Compared to several previous dust-based maps, our XPNICER maps offer higher spatial resolution and better traces regions with relatively high dust extinction than extinction maps based on optical surveys like Gaia. It serves as a high-fidelity extinction-based map, providing a complementary and independent measure of dust column densities.}
\end{itemize}

Despite representing a significant advancement, our extinction map still struggles to accurately trace extremely dense dust structures due to the limited sensitivity of current surveys. In future, upcoming optical or infrared imaging surveys such as LSST or JWST will offer unprecedented deep photometry, which can be used to create more refined extinction maps using our XPNICER technique. This work is part of the PROMISE\footnote{\url{http://promise.jounikainulainen.com}} program, which aims to derive high-dynamic-range column density data for molecular clouds by combining mid-infrared extinction, far-infrared dust emission, and near-infrared extinction data.

\section*{Acknowledgements}

We acknowledge the support from the National Key R\&D Program of China with grant 2023YFA1608000. This work was supported by the National Natural Science Foundation of China (grants No. 12473026, 12073079). This project has received funding from the European Union’s Horizon 2020 research and innovation programme under grant agreement No. 639459 (PROMISE). This research made use of Astropy,\footnote{\url{http://www.astropy.org}} a community-developed core Python package for Astronomy \citep{astropy:2013,astropy:2018}. This research has made use of the VizieR catalogue access tool, CDS,
 Strasbourg, France (DOI : 10.26093/cds/vizier). The original description 
 of the VizieR service was published in 2000, A\&AS 143, 23.

\section*{Data Availability}\label{sect:product}

We have made available the extinction maps, along with corresponding uncertainty and number density maps, produced using the XPNICER technique based on the UKIDSS/GPS photometric catalogs at \url{https://doi.org/10.57760/sciencedb.12869}. These maps are provided in a multi-extension FITS file format, with detailed explanations included in the file header.




\bibliographystyle{mnras}
\bibliography{myref} 




\appendix
\section{Comparison of source number density between UKIDSS/GPS and Pan-STARRS1 survey}\label{ap1}

\begin{figure*}
    \centering
    \includegraphics[width=1.0\linewidth]{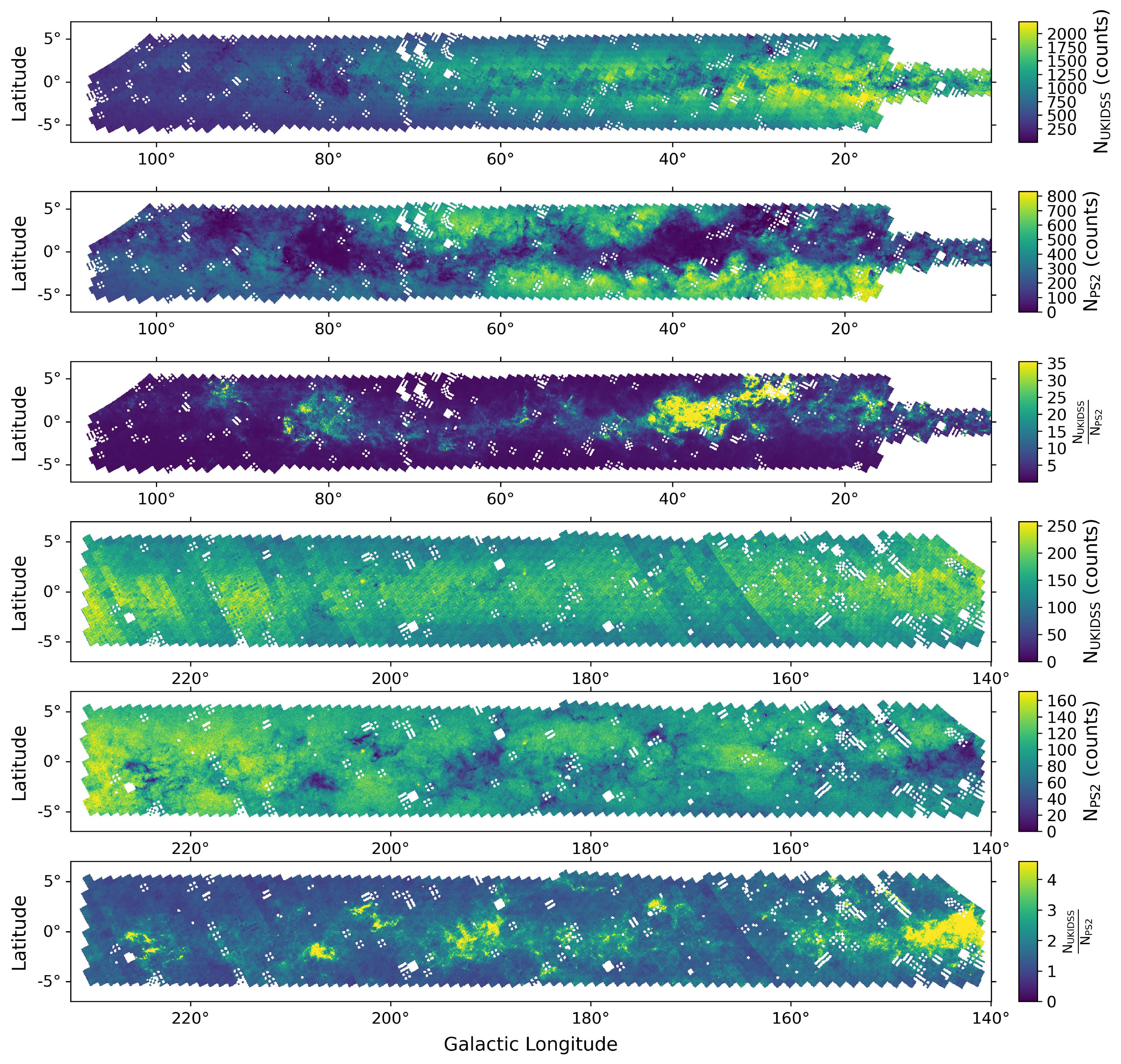}
    \caption{\revi{Point source number density maps from UKIDSS/GPS and Pan-STARRS, and their ratio map, shown over the UKIDSS/GPS survey region at a spatial resolution of 180\arcsec.}}
    \label{figap:numdens}
\end{figure*}

\revi{We extracted about 426 million NIR point sources from the UKIDSS/GPS and 2MASS catalogs. Specifically, this includes about 412 million unsaturated point sources from the UKIDSS/GPS catalog, and around 14 million bright sources from the 2MASS catalog within the UKIDSS/GPS survey region, selected with magnitude thresholds of $J <$ 13.25, $H <$ 12.75, or $K <$ 12.0 mag. Figure~\ref{figap:numdens} presents the number density maps of these $\sim$426 million NIR sources. The source density varies from a few to several hundred sources per square arcminute, with a median value of approximately 30 arcmin$^{-2}$.}

\revi{The Pan-STARRS1 (PS1) telescope, located on Mount Haleakala, Hawaii, is equipped with the Gigapixel Camera \#1 (GPC1). From 2010 to 2014, PS1 conducted the multi-epoch $3\pi$ survey, covering the northern sky ($\delta > -30^\circ$) in five broadband filters: $g_{\mathrm{P1}}$, $r_{\mathrm{P1}}$, $i_{\mathrm{P1}}$, $z_{\mathrm{P1}}$, and $y_{\mathrm{P1}}$. The single-epoch imaging reaches 5$\sigma$ depths (AB magnitudes) of 22.0, 21.8, 21.5, 20.9, and 19.7 mag in the $g_{\mathrm{P1}}$, $r_{\mathrm{P1}}$, $i_{\mathrm{P1}}$, $z_{\mathrm{P1}}$, and $y_{\mathrm{P1}}$ bands, respectively \citep{pan-starrs2016}. 
We retrieved mean photometric measurements of point sources from Data Release 2 (DR2), which were obtained by averaging detections across all epochs. The data were accessed using the MAST Pan-STARRS catalog API\footnote{\url{https://catalogs.mast.stsci.edu/docs/panstarrs.html}} through an SQL query. We selected objects with valid mean magnitudes in at least the $g_{\mathrm{P1}}$ and $r_{\mathrm{P1}}$ bands. To exclude sources affected by artifacts, we further required QfPerfect $>$ 0.85 in both $g_{\mathrm{P1}}$ and $r_{\mathrm{P1}}$. Finally, point sources were identified by requiring a small difference ($<$ 0.05 mag) between the Kron and PSF magnitudes in the $r_{\mathrm{P1}}$ band, following the criteria of \citet{psdatabase}.}   

\revi{We ultimately obtained approximately 153 million point sources from the PS1 survey within the UKIDSS/GPS survey region. Figure~\ref{figap:numdens} also presents the number density maps of PS1 point sources. As evident from the comparison, the source number density in PS1 is significantly lower than that of UKIDSS/GPS, particularly in regions with high column density. The median number density of PS1 sources is about 13 arcmin$^{-2}$.
To quantify this difference, we calculated the ratio of UKIDSS/GPS source number density ($N_{\mathrm{UKIDSS}}$) to PS1 source number density ($N_{\mathrm{PS1}}$), as shown in Fig~\ref{figap:numdens}. In some dense regions, the ratio exceeds 40, with a median value of $\sim$2 across the survey area.}

\section{Comparison of different crossmatching results with and without considering proper motions}\label{ap2}

\revi{In Sect.~\ref{sect:input-catalog}, we cross-matched about 426 million UKIDSS/GPS sources with around 52 million \gaia~DR3 sources, without applying proper motion corrections, resulting in a matched sample of about 50 million sources. The UKIDSS/GPS observations were conducted between 2005 and 2013, while the \gaia~mission began its science operations in 2014. Given this time gap, a precise cross-match would ideally account for source proper motions and the reference epochs of both surveys.}

\revi{\gaia~DR3 provides proper motion measurements for all sources. Ideally, the \gaia~astrometry should be propagated individually to the epoch of each UKIDSS/GPS observation before performing the cross-match. However, the UKIDSS/GPS catalog is a merged dataset constructed from single-band detections taken at different epochs. Therefore, a proper propagation of \gaia~astrometry would require tracing back to the single-band detection tables, applying proper motion corrections to their positions, and then performing the band-merging step based on these corrected positions, following the procedure described in the UKIDSS pipeline \citep{irwin08,hambly08}. Such a detailed treatment is beyond the scope of this paper.}

\revi{\citet{scar2018} developed a sub-arcsecond crossmatching method and applied it to create a matched catalog between \gaia~DR2, the INT Photometric H$\alpha$ Survey of the Northern Galactic Plane Data Release 2 \citep[IPHAS DR2,][]{iphas2005}, and the \textit{Kepler}-INT Survey \citep[KIS,][]{kis2012}. To avoid recalculating the \gaia~astrometry for every individual source, they divided the IPHAS and KIS catalogs into monthly epoch batches. For each batch, they assumed that the observations in different bands for a given source occurred within a short time interval, allowing them to adopt the start of the single-band observations as a reference epoch. They then propagated the \gaia~positions and proper motions to the mid-point epoch of each month. Following this approach, \citet{xgaps2023} applied the same technique to cross-match \gaia~DR3 with the INT Galactic Plane Surveys \citep[IGAPS,][]{igaps2020} and UKIDSS/GPS. The resulting catalog, called XGAPS, contains about 33 million \gaia~DR3 sources with IGAPS photometry, of which approximately 20 million also have UKIDSS/GPS measurements.}

\revi{We downloaded the XGAPS catalog from ViZieR\footnote{\url{https://cdsarc.cds.unistra.fr/viz-bin/cat/J/MNRAS/518/3137}}, which provides precise cross-matches between \gaia~DR3 and UKIDSS/GPS sources for about 33 million \gaia~DR3 objects. Independently, we performed a direct cross-match of these same \gaia~DR3 sources with the UKIDSS/GPS catalog using a matching tolerance of 0.5\arcsec~as described in Sect.~\ref{sect:input-catalog} (hereafter referred to as Xmatch). While Xmatch does not account for proper motions, XGAPS includes proper motion corrections to improve matching accuracy. We then compared the results from these two cross-matching methods.}

\begin{figure*}
    \centering
    \includegraphics[width=1.0\linewidth]{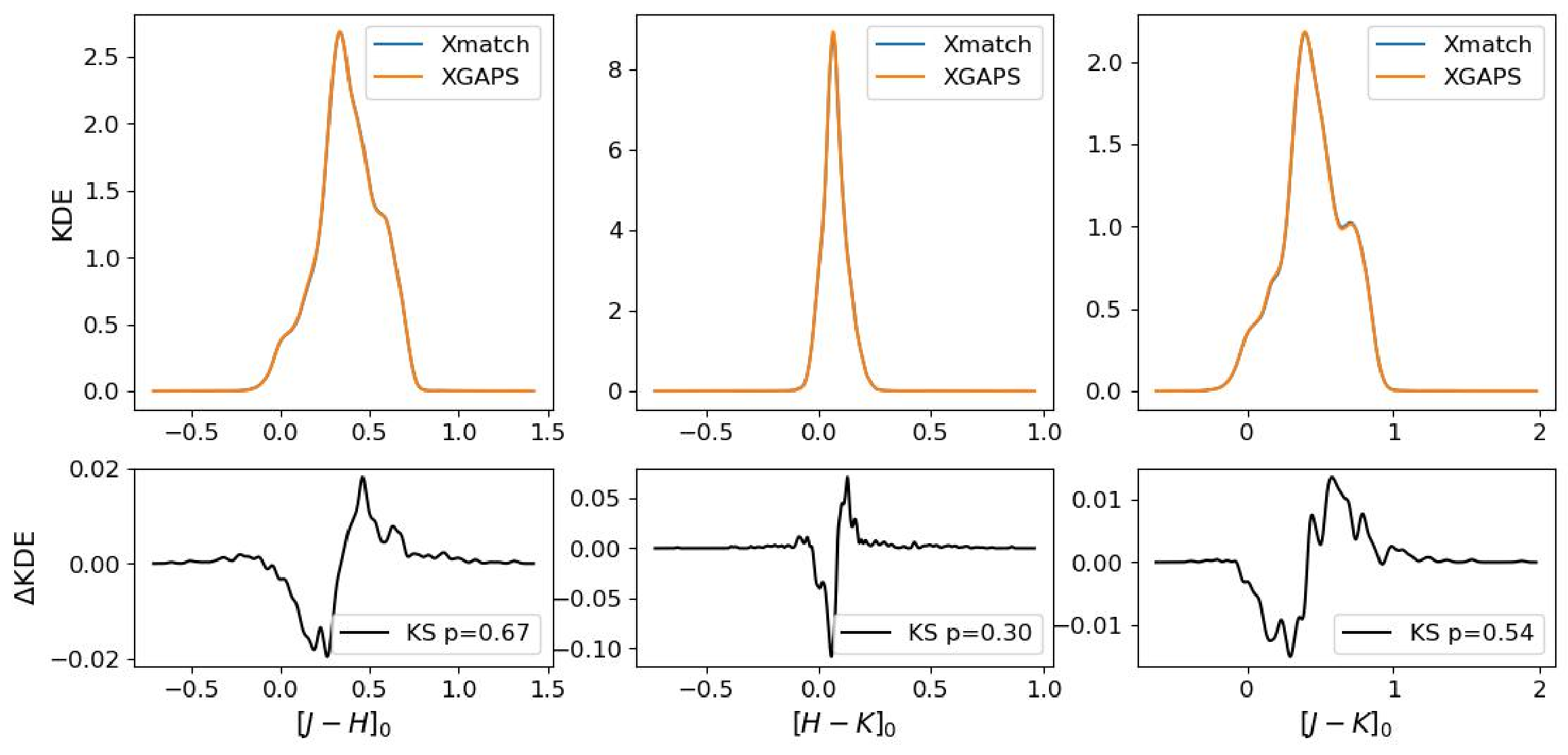}
    \caption{\revi{KDEs (\textit{top}) and KDE difference (\textit{bottom}) of the intrinsic colors of $J-H$ (\textit{left}), $H-K$ (\textit{middle}), and $J-K$ (\textit{right}) from Xmatch and XGAPS cross-matched sources, respectively.}}
    \label{fig:crossmatch}
\end{figure*}

\revi{We randomly selected 100,000 sources from the $\sim$33 million \gaia~DR3 objects, among which about 76,000 have counterparts in the UKIDSS/GPS catalog based on our Xmatch cross-match. Of these $\sim$76,000 sources, roughly 72,000 also have counterparts in the UKIDSS/GPS catalog according to the XGAPS cross-match. This implies that about 5\% of matches are mismatches when proper motions are not taken into account in the cross-matching process.
Since the Xmatch catalog is used to statistically estimate the intrinsic colors of UKIDSS/GPS sources, the presence of these mismatches should not significantly impact extinction estimates if the intrinsic color distributions from Xmatch and XGAPS samples are statistically similar. Figure~\ref{fig:crossmatch} presents the Kernel Density Estimations (KDEs) of the intrinsic colors $J-H$, $H-K$, and $J-K$ for the $\sim$76,000 Xmatch-matched sources and the $\sim$72,000 XGAPS-matched sources. The KDEs show only minor differences between the two samples. Supporting this, two-sample Kolmogorov-Smirnov (KS) tests yield $p$-values greater than 0.05 for all intrinsic colors, indicating that the null hypothesis—that the two samples come from the same distribution—cannot be rejected.
Repeating this process 1,000 times, we consistently find a mismatch contamination rate of 5–6\% when proper motions are ignored during the cross-match between \gaia~DR3 and UKIDSS/GPS. However, the intrinsic NIR colors of sources matched without proper motion corrections remain statistically indistinguishable from those matched with proper motion considered.}

\revi{Overall, directly cross-matching \gaia~DR3 and UKIDSS/GPS without accounting for proper motions results in approximately 5–6\% mismatches. However, these mismatches do not significantly affect the statistical distribution of the intrinsic NIR colors of the matched sources. Therefore, from a statistical view, neglecting proper motion corrections in the cross-match does not significantly affect our extinction estimates for UKIDSS/GPS sources.}

\section{Close view of our XPNICER extinction map}

\begin{figure*}
    \centering
    \includegraphics[width=1.0\linewidth]{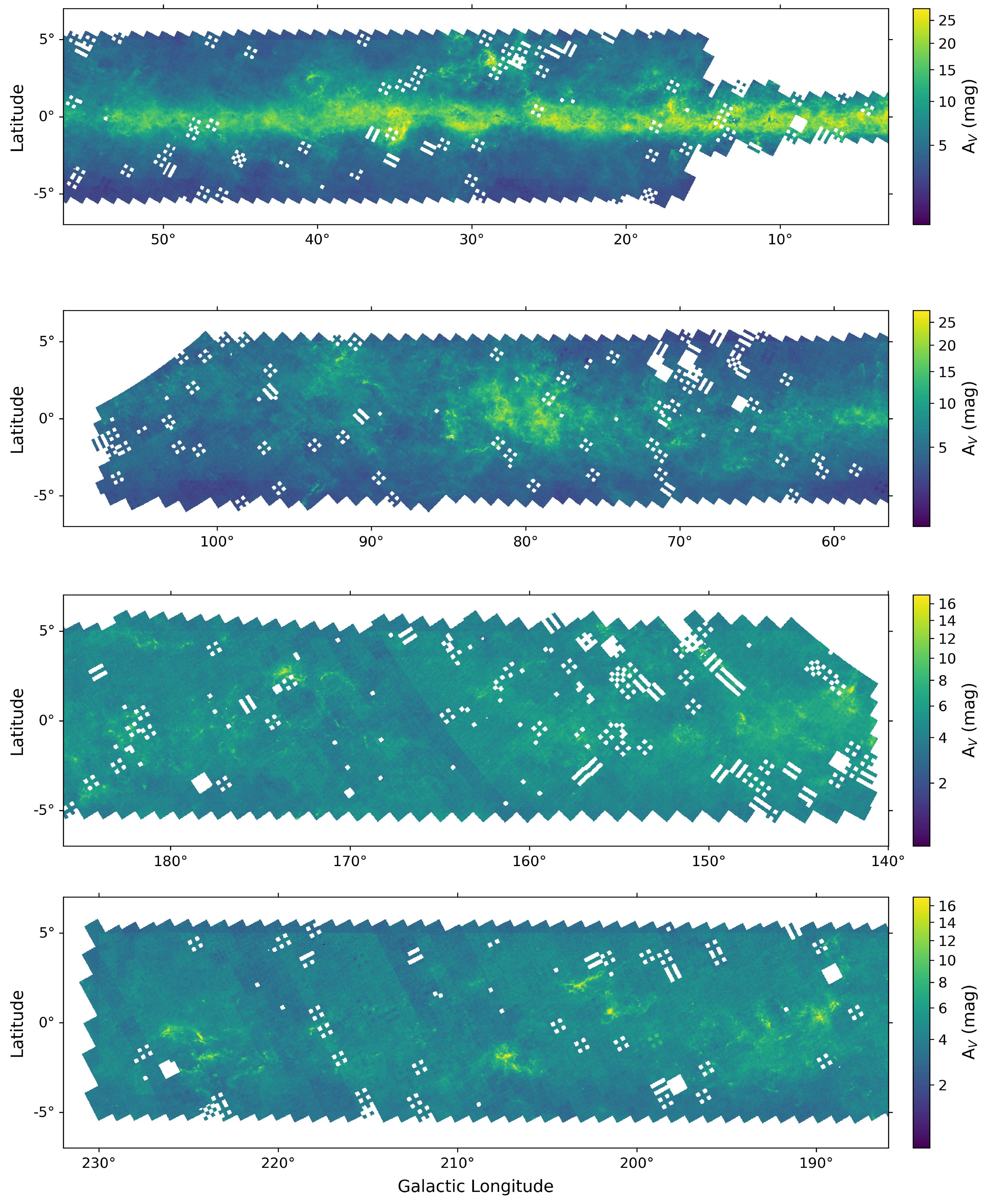}
    \caption{The XPNICER extinction maps of background sources obtained using $X_0=$~80\% and $X_1=$~95\% with the spatial resolution of \revi{90\arcsec}~for the Galactic plane area covered by UKIDSS/GPS. \revi{The zero-point offset has been subtracted from the extinction map.}}
    \label{fig:zoomextmap_disk}
\end{figure*}

\section{Tile patterns in the maps}\label{ap:tile-pattern}

\begin{figure*}
    \centering
    \includegraphics[width=1.0\linewidth]{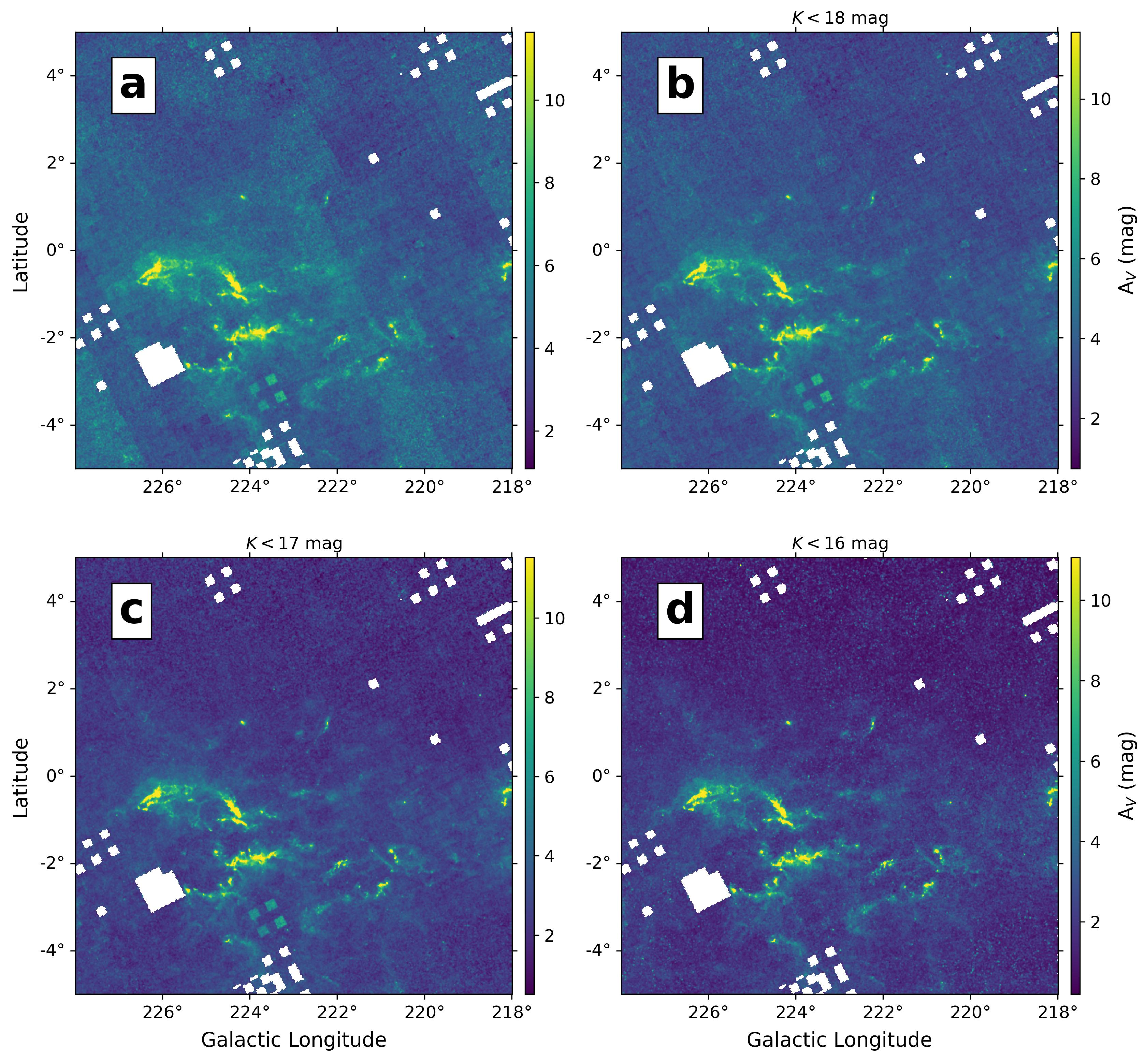}
    \caption{\revi{The XPNICER extinction maps ($A_V^{90}(X_0=80)$) produced with all background sources (\textit{a}); background sources with $K<$~18\,mag (\textit{b}); background sources with $K<$~17\,mag (\textit{c}); and background sources with $K<$~16\,mag (\textit{d}) for a 10\degr$\times$10\degr~region located at $l=$~223\degr~and $b=$~0\degr. The zero-point offset has not been subtracted from these maps.}}
    \label{fig:tile-pattern}
\end{figure*}

\revi{Tile patterns are visible in $A_V^{90}(X_0=80)$, $\delta A_V^{90}(X_0=80)$, and $N_{\mathrm{bg}}^{90}(X_0=80)$, as shown in Fig.\ref{fig:extmap_disk}. These patterns are particularly noticeable in the outer Galactic plane, for example, in several curved grid-like structures around $l \sim 160^\circ$–$170^\circ$ and $l \sim 200^\circ$–$230^\circ$. Figure\ref{fig:tile-pattern}a presents a $10^\circ \times 10^\circ$ section of $A_V^{90}(X_0=80)$, where inclined dark tile patterns can be clearly seen.}

\revi{As noted by \citet{vvvextmap}, these tile patterns arise from varying sensitivity across the UKIDSS/GPS survey. Due to the different characteristics of detectors and atmospheric conditions, the images observed at different time usually have different sensitivity. This variation occurs on scales corresponding to the field of view or individual detectors, producing the observed tile patterns.}

\revi{Such patterns can be removed by applying a uniform sensitivity cut, for example, a brightness threshold in the $K$ band. Figures~\ref{fig:tile-pattern}b, c, and d show extinction maps generated using background sources with $K < 18$, $K < 17$, and $K < 16$ mag, respectively. While some residual tile patterns remain in panels b and c, the patterns nearly disappear at $K < 16$ mag.}

\revi{However, this threshold approaches the sensitivity limit of the 2MASS survey, which is much brighter than that of UKIDSS/GPS ($K\sim18$ mag). Consequently, applying such a magnitude cut reduces the number of background sources, resulting in a significant loss of the resolution and dynamic range of the extinction maps. Therefore, we chose not to correct for tile patterns in this paper.}

\bsp	
\label{lastpage}
\end{document}